\documentclass[12pt,preprint]{aastex}

\def\lesssim{\mathrel{\hbox{\rlap{\hbox{\lower4pt\hbox{$\sim$}}}\hbox{$<$}}}}
\def\gtrsim{\mathrel{\hbox{\rlap{\hbox{\lower4pt\hbox{$\sim$}}}\hbox{$>$}}}}
\newcommand{\etal }{{et al.}} 
\newcommand{\msun}{\thinspace M_\odot} 
\newcommand{\vect}[1]{\mbox{\boldmath$#1$}}
\newcommand{\cm  }{\,{\rm cm}^{-3} } 
\newcommand{\ob}{\varepsilon_{\rm ob}}
\newcommand{\ar}{\varepsilon_{\rm ar}}
\newcommand{\bw}{$B$-$\Omega$ }
\newcommand{\rhoc}{\rho_{\rm c}}
\newcommand{\dfrac}[2]{{\displaystyle \frac{#1}{#2}}  }

\shorttitle{Cloud Evolution with Oblique Field}
\shortauthors{Machida \etal 2005}

\begin{document}

\title{Evolution of Rotating Molecular Cloud Core with Oblique Magnetic Field}

\author{Masahiro N. Machida\altaffilmark{1} } 
\affil{Department of Physics, Graduate School of Science, Kyoto University, Sakyo-ku, Kyoto 606-8502, Japan}

\author{Tomoaki Matsumoto\altaffilmark{2}}
\affil{Faculty of Humanity and Environment, Hosei University, Fujimi, Chiyoda-ku, Tokyo 102-8160, Japan}

\author{Tomoyuki Hanawa\altaffilmark{3}}
\affil{Center for Frontier Science, 
Chiba University, Yayoicho 1-33, Inageku, Chiba 263-8522, Japan}

\author{Kohji Tomisaka\altaffilmark{4}} 
\affil{National Astronomical Observatory, Mitaka, Tokyo 181-8588, Japan}

\altaffiltext{1}{machidam@scphys.kyoto-u.ac.jp}
\altaffiltext{2}{matsu@i.hosei.ac.jp}
\altaffiltext{3}{hanawa@cfs.chiba-u.ac.jp}
\altaffiltext{4}{tomisaka@th.nao.ac.jp; also School of Physical Sciences, Graduate University of Advanced Study (SOKENDAI)}

\begin{abstract}
We studied the collapse of rotating molecular cloud cores with inclined magnetic fields, based on three-dimensional numerical simulations.
The numerical simulations start from a rotating Bonnor--Ebert isothermal cloud in a uniform magnetic field.  The magnetic field is initially taken to be inclined from the rotation axis.  
As the cloud collapses, the magnetic field and rotation axis change their directions.  
When the rotation is slow and the magnetic field is relatively strong, the direction of the rotation axis changes to align with the magnetic field, as shown earlier by Matsumoto \& Tomisaka. When the magnetic field is weak and the rotation is relatively fast, the magnetic field inclines to become perpendicular to the rotation axis.  In other words, the evolution
of the magnetic field and rotation axis depends on the relative strength of the rotation and magnetic field.  
Magnetic braking acts to align the rotation axis and magnetic field, while the rotation causes the magnetic field to incline through dynamo action.  
The latter effect dominates the former when the ratio of the angular velocity to the magnetic field is larger than a critical value $ \Omega _0/ B _0 \, > \, 0.39\, G^{1/2} \, c_s^{-1} $, where $ B _0 $, $ \Omega _0 $, $ G $, and $ c _s^{-1}$ denote the initial magnetic field, initial angular velocity, gravitational constant, and sound speed, respectively.
When the rotation is relatively strong, the collapsing cloud forms a disk perpendicular to the rotation axis and the magnetic field becomes nearly parallel to the disk surface in the high density region.  
A spiral structure appears due to the rotation and the wound-up magnetic field in the disk. 

\end{abstract}


\section{Introduction}
   Magnetic fields and rotation are believed to play important roles in the gravitational collapse of molecular cloud cores.
   For example, the outflow associated with a young star is believed to be related to the magnetic field and rotation of a protostar and its disk.
   The rotation of various molecular clouds have been studied by \citet{caselli02}, \citet{goodman93}, and \citet{arquilla86}, who found that the rotation energy is not negligible compared with the gravitational energy. 
  \citet{crutcher99} obtained strengths of the magnetic field for various molecular clouds by Zeeman splitting observations and also concluded that the magnetic energy of a molecular cloud is comparable to the gravitational energy. 
   The direction of the magnetic field is also crucial for cloud evolution, because it controls the direction of outflow and the orientation of the disk.
   Polarization observations of young stellar objects suggest that circumstellar dust disks around young stars are aligned perpendicular to the magnetic field \citep[e.g.,][]{moneti84,tamura89}.  
  However, the direction of a large-scale magnetic field of an ambient cloud and that of a small-scale magnetic field around a molecular core do not always coincide.
  For example, the Barnard 1 cloud in Perseus exhibits field directions different from the ambient field in three of its four cores \citep{matthews02}.
 Recently, high-resolution observations by \citet{matthews05} have shown that the polarization angle measured in the OMC-3 region of the Orion A cloud changes systematically across the core.
 These findings indicate that the spatial configuration of magnetic field lines in a molecular core is not simple.

\citet{dorfi82,dorfi89} and \citet[][hereafter MT04]{matsumoto04} numerically investigated the contraction of a molecular cloud when the magnetic field lines are not parallel to the rotation axis (non-aligned rotator).
  \citet{dorfi82,dorfi89} showed that a formed disk changes its shape into a bar or a ring depending on the angle between the magnetic field and the rotation axis at the initial stage.  
 However, the evolution was not calculated to high densities.
 MT04 reinvestigated the collapse of rotating molecular cloud cores threaded by oblique magnetic fields by high-resolution numerical simulations.
 They found that strong magnetic braking associated with outflow causes the direction of the angular momentum to converge to that of the local magnetic field, resulting in convergence of the local magnetic field, angular momentum, outflow, and disk orientation.
 However, their study was restricted to clouds with an intermediate rotation rate ($\Omega = 7.11 \times 10^{-7}$ yr$^{-1}$).
It is expected that the evolution is affected not only by the magnetic field strength but also by the rotation rate.
In this paper, we extend the parameter range of the MT04 study and explore the evolution of rotating magnetized clouds more generally.

 In the case of a cloud in which the magnetic field is parallel to the rotation axis (aligned rotator), four distinct evolutions are observed according to the magnetic field strength and rotation rate \citep[][hereafter Papers I, II, and III]{machida04,machida05a,machida05b}.
 In the isothermal regime a contracting disk is formed perpendicular to the magnetic field and the rotation axis.
 In the disk, the magnetic field strength, angular rotation speed, and gas density are correlated with one another and satisfy the magnetic flux-spin relation:
\begin{equation}
   \frac{\Omega_{zc}^2}{(0.2)^2 \times 4 \pi G \rho_c} + \frac{B_{\rm zc}^2}{(0.36)^2 \times 8 \pi c_s^2 \rho_c} \equiv F (\Omega_{\rm zc}, B_{\rm zc}, \rho_c) \simeq 1,
\label{eq:UL}
\end{equation}
where $\Omega_{zc}$, $B_{zc}$, $\rhoc$, $c_s$, and $G$ are the angular velocity, magnetic flux density, gas density at the center, isothermal sound speed, and gravitational constant, respectively.

In the case of a weak magnetic field and a small rotation rate, $F < 1$, the cloud contracts spherically (spherical collapse).
 This increases $F$ and a self-similarly contracting disk forms when $F\simeq 1$ is reached.
On the other hand, a cloud with a strong magnetic field and/or a fast rotation rate, $F > 1$, contracts only in the direction of the magnetic field and rotation axis (vertical collapse).
 This reduces $F$ and  a self-similarly contracting disk forms when $F \simeq 1$.
 Hence, $F$ controls the mode of contraction (spherical or vertical collapse).
This is understood as follows: for $F < 1$ the support forces are deficient (support-deficient model), while for $F > 1$ the cloud is supported laterally by rotation and/or the magnetic field (support-sufficient model).

 The evolution can be further divided into two categories depending on whether the magnetic or centrifugal forces are more effective in forming the disk.
A model with a spin-to-magnetic flux ratio $\Omega_0/B_0$ larger than a critical value $(\Omega/B)_{\rm crit} = 0.39 G^{1/2} c_s^{-1} $ forms a disk mainly supported by the centrifugal force, while that with a smaller ratio forms a disk by the Lorentz force.

 In addition to the differences in the evolution of the pre-protostellar phase (isothermal phase), subsequent evolution is affected by the magnetic field and rotation.
 A first core consisting of adiabatic H$_2$ molecular gas experiences fragmentation only if the initial cloud is rotation dominated,  $(\Omega_0/B_0) > (\Omega/B)_{\rm crit}$.  
In the support-deficient regime ($F<1$), fragmentation proceeds through a deformation forming a ring, while in the support-sufficient regime ($F > 1$) fragmentation from a bar appears as well as ring fragmentation.

Based upon our previous results (Papers II and III),  all models of \citet{dorfi82,dorfi89} and MT04 belong to a type of ``support-sufficient'', $F > 1$, and ``magnetic-force-dominant'', $\Omega_0/B_0 < (\Omega/B)_{\rm crit}$, models.
 In this paper, we investigate the evolution of a rotating isothermal cloud with an inclined magnetic field, not restricted to the above type, 
in large parameter space, and also explore three other models which have not been previously studied. 

 The plan of the paper is as follows. The numerical method of our computations and the framework of our models are given in \S 2 and 
 the numerical results are presented in \S 3.  
  We discuss the geometry of the collapse in \S 4, and compare our results with previous 
works and observations in \S 5.

\section{Numerical Model}
 Our initial settings are almost the same as those of MT04.
 To study the cloud evolution and disk formation, we use the three-dimensional magnetohydrodynamical (MHD) nested grid method.
 We assume ideal MHD equations including the self-gravity:  
\begin{eqnarray} 
& \dfrac{\partial \rho}{\partial t}  + \nabla \cdot (\rho \vect{v}) = 0, & \\
& \rho \dfrac{\partial \vect{v}}{\partial t} 
    + \rho(\vect{v} \cdot \nabla)\vect{v} =
    - \nabla P - \dfrac{1}{4 \pi} \vect{B} \times (\nabla \times \vect{B})
    - \rho \nabla \phi, & \\ 
& \dfrac{\partial \vect{B}}{\partial t} = 
   \nabla \times (\vect{v} \times \vect{B}), & \\
& \nabla^2 \phi = 4 \pi G \rho, &
\end{eqnarray}
 where $\rho$, $\vect{v}$, $P$, $\vect{B} $, and $\phi$ denote the density, 
velocity, pressure, magnetic flux density, and gravitational potential, 
respectively. 
 The gas pressure is assumed to be expressed by the gas density (barotropic gas) as
\begin{equation}
P = c_s^2 \rho \left[
1 + \left( \dfrac{\rho}{\rho_{\rm cri}}   \right)^{2/5} 
\right],
\label{eq:eos}
\end{equation}
where $c_s = 190\, {\rm m}\, {\rm s^{-1}}$ and $ \rho_{\rm cri} = 1.9205 \times 10^{-13} \, \rm{g} \, \cm$  ($n_{\rm cri} = 5\times 10^{10} \cm$ for an assumed mean molecular weight of 2.3).
 This equation of state implies that the gas is isothermal at $T = 10$ K for $n \ll n_{\rm cri}$ and is adiabatic for $n \gg n_{\rm cri} $ \citep{masunaga00}. 
 For convenience, we define the core formation epoch as that for which the central density ($n_c$) exceeds  $n_{\rm cri} $.
 We also call the period for which $n_c < n_{\rm cri}$ the isothermal phase, and the period for which $n_c \ge n_{\rm cri}$ the adiabatic phase.

 In this paper, a spherical cloud with a critical Bonnor--Ebert \citep{ebert55, bonnor56} density profile, $\rho_{\rm BE}$, is assumed as the initial condition, although a filamentary cloud is assumed in Papers I--III. 
 The cloud rotates rigidly ($\Omega_0$) around the $z$-axis and has a uniform magnetic field ($B_0$).
 To promote contraction, we increase the density by a factor $f$ (density enhancement factor) as 
\begin{eqnarray}
\rho(r) = \left\{
\begin{array}{ll}
\rho_{\rm BE}(r) \, f & \mbox{for} \; \; r < R_{c}, \\
\rho_{\rm BE}(R_c)\, f & \mbox{for}\; \;  r \ge R_{c}, \\
\end{array}
\right. 
\end{eqnarray}
 where $r$ and $R_c$ denote the radius and the critical radius for a Bonnor--Ebert sphere, respectively.
 We assume $\rho_{\rm BE}(0) = 7.39 \times 10^{-19}\, {\rm g} \,\cm$, which corresponds to a central number density of $n_{c,0} = 5\times 10^4\cm$.
Thus, the critical radius for a Bonnor--Ebert sphere $R_c = 6.45\, c_s [4\pi G \rho_{BE}(0)]^{-1/2}$ corresponds to $ R_c = 2.05 \times 10^4$ AU for our  settings. 

The initial model is characterized by three nondimensional parameters: $\alpha$, $\omega$, and $\theta_0$.
The magnetic field strength and rotation rate are scaled using a central density $\rho_0 = \rho_{\rm BE}(0) f$ as
\begin{equation}
\alpha =  B_0^2 / (4\pi \, \rho_0 \, c_{s}^2),
\label{eq:alpha}
\end{equation}
and
\begin{equation}
\omega = \Omega_0/(4 \pi\,  G \, \rho_0  )^{1/2}.
\label{eq:omega}
\end{equation}
The parameter $\theta_0$ represents the angle between the magnetic field and the rotation axis ($z$-axis).
 Thus, the initial magnetic field is given by
\begin{eqnarray}
\left(
\begin{array}{l}
B_x \\
B_y \\
B_z \\
\end{array}
\right)
= B_0
\left(
\begin{array}{l}
\mbox{sin}\, \theta_0 \\
0 \\
\mbox{cos}\, \theta_0 \\
\end{array}
\right)
\end{eqnarray}
in Cartesian coordinates ($x$, $y$, $z$).
The above definitions of $\alpha$ and $\omega$ are the same as those of Papers I--III.
Although we added a finite non-axisymmetric perturbation in Papers I--III, we assume no explicit non-axisymmetric perturbation in this paper.

 We calculated 36 models, widely covering the parameter space.
 20 typical models are listed in Table~\ref{table:init}.
 The model parameters ($\alpha$, $\omega$, and $\theta_0$); density enhancement factor ($f$); ratio of the thermal ($\alpha_0$),  rotational ($\beta_0$),  and magnetic ($\gamma_0$) energies to the gravitational energy;\footnote{
Denoting the thermal, rotational, magnetic, and gravitational energies as $U$, $K$, $M$, and $W$, the relative factors against the gravitational energy are defined as $\alpha_0 = U/|W|$, $\beta_0 = K/|W|$, and $\gamma_0 = M/|W|$.
}
initial central number density ($n_{0}$),  magnetic field strength ($B_0$); angular velocity ($\Omega_0$); and  total mass inside the critical radius ($r<R_{\rm c}$) are summarized in this table.
 The models SF, MF, and WF in MT04 are also listed.
 The clouds in MT04 have stronger (models SF and MF) or equivalent (model WF) magnetic fields compared to our models and have an intermediate rotation rate.

We calculated cloud evolutions up to $n \simeq 10^{14} \cm$\ using an ideal MHD approximation.
The ideal MHD approximation is fairly good as long as the gas density is lower than $ \sim 10 ^{12} $~cm$^{-3}$ \citep{nakano76,nakano02}.  
However, ohmic dissipation affects protostellar collapse, especially at high densities exceeding $ n \simeq 10^{12} \cm$ \citep{nakano02}.
\citet{nakano02} have shown that the magnetic field is coupled with gas for $n \la 10^{12} \cm$, indicating that the assumption of an ideal MHD is valid in the isothermal phase.  
In the adiabatic phase, the number density in the adiabatic core exceeds $\sim 10^{12} \cm$, and the magnetic field begins to decouple as the ohmic dissipation being effective. 
Our simulation may therefore overestimate the magnetic field, especially for a dense core ($n \gtrsim 10^{12} \cm$). 
Since we are interested in the direction of the magnetic field, rotation axis, and disk normal in the isothermal phase, we show the results of cloud evolution mainly for low density cores ($n \lesssim 10^{12} \cm$) in which the ideal MHD approximation is valid.  
The effect of ohmic dissipation and cloud evolution, outflow, and jets in high density cores will be investigated in a subsequent paper.

We adopt the nested grid method (for details, see Paper II).
 The nested grid consists of concentric hierarchical rectangular grids
to give a high spatial resolution near the origin.
  Each level of the rectangular grid has the same number of cells ($128 \times 128 \times 128 $), but the cell width $h(l)$ depends on the grid level $l$.
 The cell width is reduced by a factor of 1/2 as the grid level increases by 1 ($l \rightarrow l+1$).
 We begin our calculations with 6 grid levels ($l=1$--6).
 The box size of the initial finest grid $l=6$ was chosen to be $2 R_{\rm c}$. 
 The coarsest grid ($l=1$), therefore, has a box size equal to $2^6\, R_{\rm c}$. 
 A boundary condition is imposed at $r=2^6\, R_{\rm c}$, where the magnetic field and ambient gas rotate at an angular velocity of $\Omega_0$ (for details see MT04).
 Due to this large simulation box, it takes $t\simeq 40$ free-fall time until the Alfv\'en wave generated at the cloud center reaches the simulation boundary, even in the model with the strongest magnetic field.
 Hence, the boundary condition does not affect the central cloud because our calculations end within $\simeq 10$ free-fall time.
  The highest level of grids changes dynamically.
  A new finer grid was generated whenever the minimum local Jeans length 
$ \lambda _{\rm J} $ become smaller than $ 8\, h (l{\rm max}) $, where $h$ is the cell width.
 The maximum level of grids was restricted to $l_{\rm max} = 20$ in typical models.
 Since the density is highest in the finest grid, the generation of new grid ensures the Jeans condition of \citet{truelove97} with a margin of a factor of 2.
 We adopted the hyperbolic divergence $\vect{B}$ cleaning method
of \citet{dedner02}.

\section{Results}
   Our models are characterized mainly by the strength of the magnetic field ($\alpha$) and the angular velocity ($\omega$). We calculated six groups of models, groups A--F, distinguished by the values of $\alpha$ and $\omega$ (see Table~\ref{table:init}).
   The models are designated by group (A, B, C, D, E, or F) and $\theta_0$ (0$\degr$, 30$\degr$, 45$\degr$, or 60$\degr$). 
Hence, model A00 has $\alpha = 0.01$, $\omega =0.01$, and $\theta_0 = 0\degr$, while model A45 has the same $\alpha$ and $\omega$, but $\theta_0$ = 45$\degr$.
  Models with $\theta_0 = 0 \degr$ are ``aligned rotators'' and those with  $\theta_0 \ne 0 \degr$ are ``non-aligned rotators.''
  The aligned rotator models of groups A, B, C, and D (A00, B00, C00, and D00) have the same $\alpha$ and $\omega$ as groups A, B, C, and D in Papers II and III 
and have the same magnetic field strength and angular velocity at the center, although the distributions of the density, magnetic field, and angular velocity are different.
Groups E and F are newly added in this paper.

The evolution of the aligned rotator models can be divided into four types (\S 1) by two criteria: $F \lessgtr 1$ and $\Omega_0/B_0 \lessgtr 0.39 G^{1/2} c_s^{-1}$.
 Groups A, B, and F are magnetic-force-dominant, $\Omega_0/B_0 < 0.39 G^{1/2} c_s^{-1}$ for $\theta = 0$, and Groups C, D, and E are rotation-dominant.
 The models are described in the following subsections (\S 3.1--\S 3.3).

\subsection{Magnetic-Force-Dominant Disks}
 In this subsection, we show the evolutions of groups A, B, and F.
 These groups are expected to form magnetic-force-dominant disks in the isothermal phase since $\Omega_0/\vect{B_0} < 0.39 G^{1/2} c_s^{-1}$.
 Before disk formation, the cloud is expected to collapse spherically in groups A and B, since they are support-deficient models, $F \lesssim 1$. 
 On the other hand, one-dimensional collapse along the magnetic field lines is expected for model F, since this is a support-sufficient model, $F \gtrsim 1$.

\subsubsection{Support-deficient models}
\begin{figure}  
\plotone{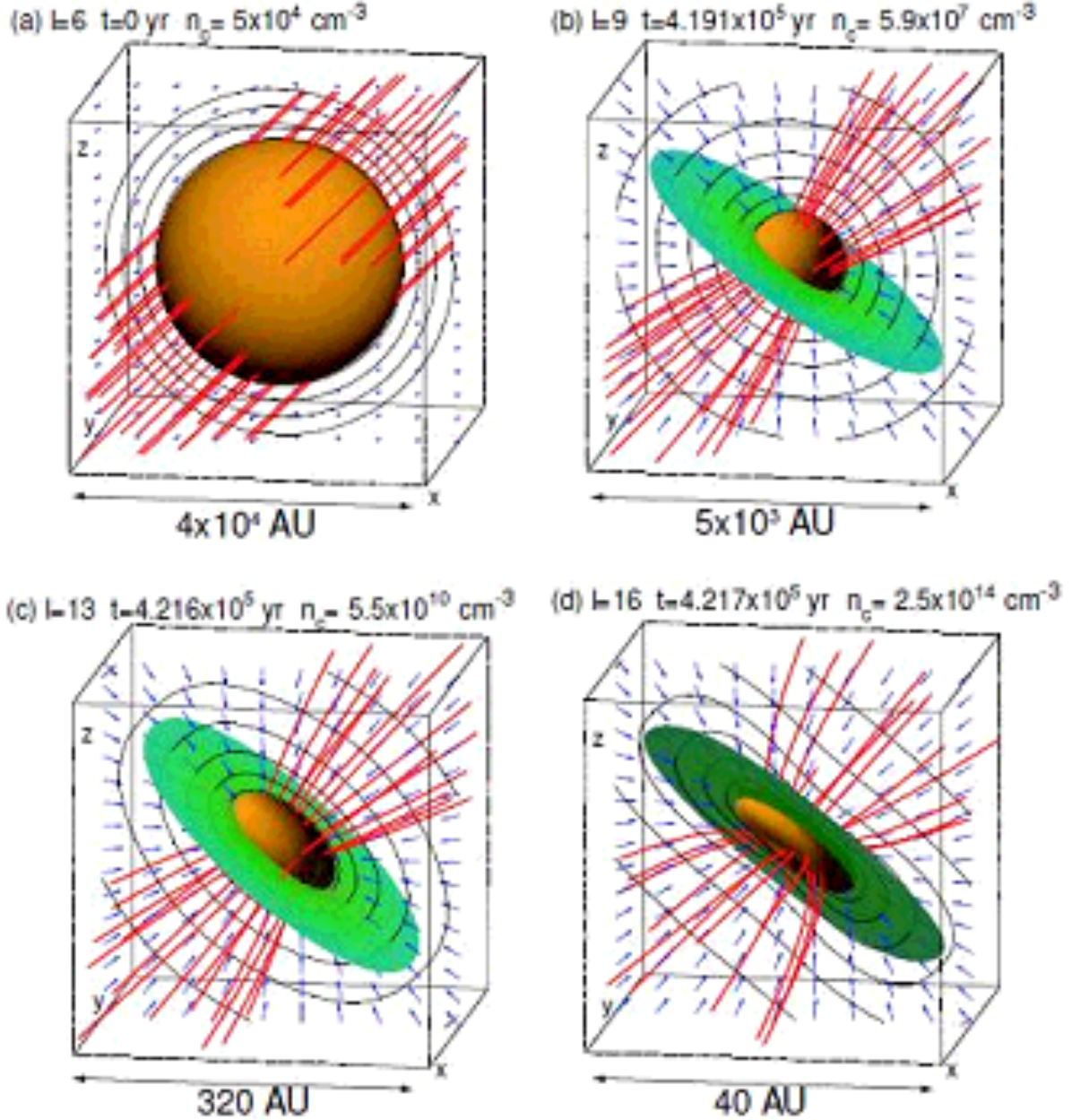}
\caption{Time sequence of model A45.
The structure of the high-density region ($n > 0.1 n_c$\,; isosurface), density contours (contour lines), velocity vectors (arrows), and magnetic field lines (streamlines) are plotted in panels  (a)--(d).
 The green colored disk indicates the region of $n > 1/100\,n_c$ on the mid-plane parallel to the disk-like structure (perpendicular to the disk normal). 
The panels show the stages (a) $n_c = 5\times 10^4\cm$ (initial stage), (b) $n_c = 5.9 \times 10^7 \cm$, (c) $n_c = 5.5 \times 10^{10} \cm$, and (d) $n_c = 2.5 \times 10^{14} \cm$.
The level of the finest grid ($l$), elapsed time from the beginning ($t$), and central number density ($n_c$) are listed at the top of each panel.
The size of the grid is also shown.
}
\label{fig:1}

\end{figure}

 In this section, we consider the evolution of groups A and B.
 Figures~\ref{fig:1} (a)--(d) shows the evolution for model A45.
 Model A45 has the parameters $\alpha=0.01$, $\omega=0.01$, and $\theta_0=45\degr$.
 This cloud rotates slowly ($\omega = 0.01$) around the $z$-axis, and has a weak magnetic field ($\alpha = 0.01$).
 Figure~\ref{fig:1} (a) shows the initial state of model A45.
 The cloud is threaded by a magnetic field running in the direction $\theta_0=45\degr$.
 Figure~\ref{fig:1} (b) shows the structure when the central density reaches $n_c = 5.9 \times 10^7\cm$.
 Figure~\ref{fig:1} (b) ($l=9$) covers $(1/8)^3$ of the volume of Figure~\ref{fig:1} (a) ($l=6$).
 Figure~\ref{fig:1} (b) shows that the magnetic field lines run in a direction at an angle $\theta_B \simeq 45\degr$ from the $z$-axis, similar to the large-scale view ($l=6$) of Figure~\ref{fig:1} (a), though they are squeezed near the center.
 The green colored disks in Figures~\ref{fig:1} (b), (c), and (d) indicate regions of $\rho> (1/100) \,\rhoc$ on the mid-plane parallel to the disk-like structure (perpendicular to the disk normal). 
 From the density contour lines in the $x$-$z$ plane of Figure~\ref{fig:1} (b), it can be seen that the high density region is slightly flattened and the cloud collapses along the magnetic field lines.
 We derive three principal axes ($a_1 \ge a_2 \ge a_3$) from the moment of inertia according to MT04.
 We define the shortest axis $a_3$ to be the disk normal axis and define $a_1$ and $a_2$ to be the long and  short axes on the disk mid-plane. 
 The oblateness and axis-ratio of the high-density region of $\rho \ge 0.1 \rho_c$ 
are defined here as $\ob \equiv (a_1 a_2)^{1/2}/a_3$ and  $\ar \equiv a_1/a_2-1$, respectively.

\begin{figure}   
\epsscale{0.55}
\plotone{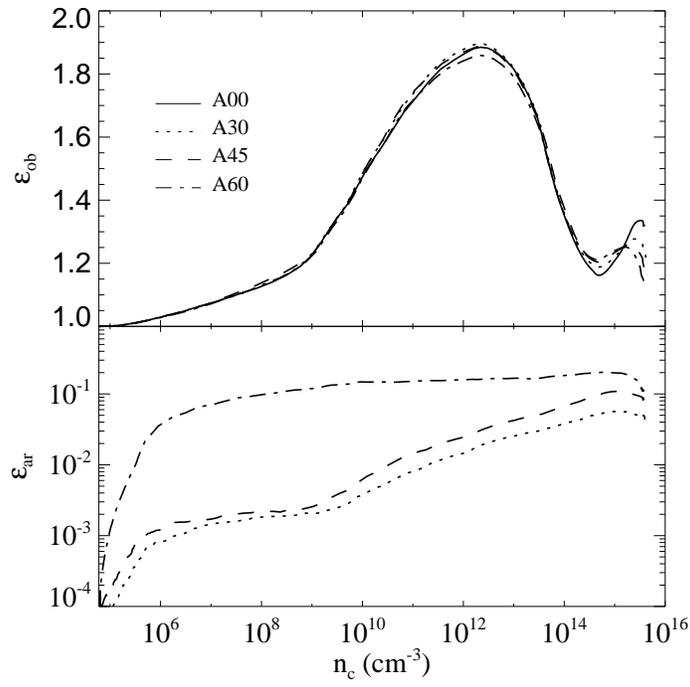}
\caption{
Oblateness (upper panel) and axis-ratio (lower panel) against central 
density for group A (models A00, A30, A45, and A60). 
}
\label{fig:2}
\end{figure}

 Figure~\ref{fig:2} plots the oblateness (upper panel) and axis-ratio (lower panel) against
 the central density for group A.
 For model A45, the oblateness reaches $\ob \simeq 1.1$ at the epoch of Figure~\ref{fig:1} (b) ($n_c = 5.9 \times 10^7 \cm$).
 This means that the high-density region maintains a spherical structure at this epoch. 
 The axis-ratio grows only to $\ar \simeq 10^{-3}$ at the same epoch.

\begin{figure}   
\epsscale{0.55}
\plotone{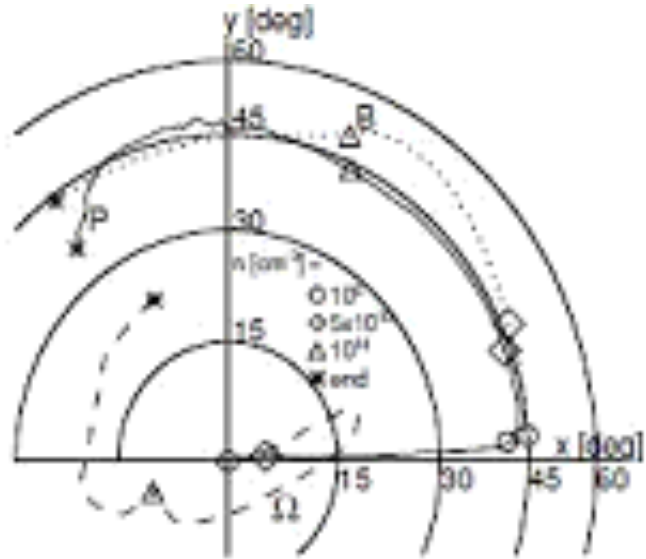}
\caption{
Loci of directions of the magnetic field $\vect{B}$ (dotted line), rotation axis $\vect{\Omega}$ (broken line), and disk normal $\vect{P}$ (solid line) for model A45.
The symbols $\circ$, $\diamond$, $\vartriangle$, and $*$ denote the stages $n_c = 10^6 \cm$, $5\times 10^{10}\cm$, $10^{14}\cm$, and the final epoch, respectively.
}
\label{fig:3}
\end{figure}

 Figure~\ref{fig:1} (c) shows the high-density region at the core formation epoch ($n_c = 5 \times 10^{10}\cm$).
  The angle between the disk normal and the $z$-axis has increased to $\theta_p \simeq 41 \degr$, although that between the magnetic field lines and the $z$-axis is $\theta_B\simeq 45\degr$,  similar to Figure~\ref{fig:1} (b).
 Figure~\ref{fig:3} shows the loci of the magnetic field $\vect{B}$ ($\theta_B, \phi_B$), the rotation axis $\vect{\Omega}$ ($\theta_{\Omega}, \phi_{\Omega}$), and the normal vector of the disk $\vect{P}$ ($\theta_{P}, \phi_{P}$), all averaged within $\rho > 0.1 \rhoc$.
 Each vector is projected on the $z=0$ plane and 
 the distance from the origin represents the angle between a given vector and the $z$-axis. 
 For example, a vector parallel to the $y$-axis is plotted at (0$\degr$, 90$\degr$).
 The dotted line in Figure~\ref{fig:3} indicates that the magnetic field rotates around the $z$-axis from $\phi_B =0\degr$ to $125 \degr$ during the calculation, keeping the same angle with respect to the  $z$-axis ($\theta_B \simeq 45 \degr$).
   It can also be seen that the disk normal (the solid line) stays parallel to the magnetic field after  $n_c > 10^6 \cm$, showing that a disk structure is formed perpendicular to the magnetic field and rotates with the magnetic field.
 On the other hand, the rotation axis, represented by the broken line, points in the direction $\theta_\Omega \lesssim 15\degr$ before $n_c \lesssim 5 \times 10^{10} \cm$.
 Thus, the rotation axis maintains its initial direction in the isothermal phase.
 We summarize the angles of the magnetic field ($\theta_B$, $\phi_B$),  rotation axis ($\theta_\Omega$, $\phi_\Omega$), and  disk normal ($\theta_P$, $\phi_P$) at the end of the isothermal phase in Table~\ref{table:results}.
 The angles between the magnetic field and the rotation axis $\psi_{B\Omega}$, those between the magnetic field and disk normal $\psi_{BP}$, and those between the rotation axis and the disk normal $\psi_{\Omega P}$ are also listed in Table~\ref{table:results}.
 For example, the angle $\psi_{B\Omega}$ is calculated as
\begin{equation}
   \psi_{B \Omega} \equiv \mbox{sin}^{-1}  \frac{|\vect{B} \times \vect{\Omega}| }{|\vect{B}|\;|\vect{\Omega}|}.
\end{equation}
 We can confirm from Table~\ref{table:results} that the magnetic field and disk normal have 
almost the same directions ($\psi_{B P} = 4 \degr$) and that the disk normal is inclined from the rotation axis at an angle of $\psi_{\Omega P}=42 \degr$ at the end of the isothermal phase.
 The oblateness reaches $\ob=1.45$ at the core formation epoch (Figure~\ref{fig:2} upper panel).
 This oblateness is smaller than that of model AS ($\ob = 2.9$) of Paper II.
 The increase in the initial central density from $5 \times 10^2 \cm$ (Paper II) to $5 \times 10^4 \cm$ suppresses the growth of oblateness.
The disk forms slowly compared with models with large $\alpha$ or $\omega$, similar to Paper II.
 The axis-ratio grows only to $\ar = 7.1 \times 10^{-3}$ at the end of the isothermal phase.

 Figure~\ref{fig:1} (d) shows the central region at 140 yr after the core formation epoch ($n_c = n_{\rm cri}$).
At this stage, the directions of the magnetic field and disk normal are parallel to each other and co-rotate around the $z$-axis, keeping the angles $\theta_B, \theta_{P} \simeq 45 \degr.$ 
 The magnetic field and disk normal vector continue to rotate up to $\phi_B, \phi_P \simeq 125\degr$ at the end of the calculation period ($n_c \sim 10^{14} \cm$).
 The rotation axis begins to move away from the $z$-axis at the beginning of the adiabatic phase, and has an angle $\theta_\Omega \simeq 20 \degr$ with respect to the $z$-axis at the end of the calculation (asterisk). 
 It can be seen that the three axes ($\vect{B}$, $\vect{\Omega}$, and $\vect{P}$) are in alignment in the adiabatic stage, as seen in MT04.
 Although the disk becomes thinner in the adiabatic phase, the oblateness is no more than  $\ob \simeq 2$, as shown in Figure~\ref{fig:2}.
 When the central density exceeds $n_c \simeq 10^{12} \cm$, the oblateness gradually decreases  and the central region becomes spherical owing to the thermal pressure.
 Figure~\ref{fig:2} indicates that the oblateness depends only slightly on  $\theta_0$.
 On the other hand, the evolution of the axis-ratio clearly depends on  $\theta_0$.
 The axis-ratio grows faster in models with larger $\theta_0$.
The growth of the axis-ratio must be due to non-axisymmetry in the plane perpendicular to the magnetic field in magnetic-force-dominated models.
 Such non-axisymmetry is induced by the centrifugal force due to the rotation motion with $\vect{\Omega}$ in the disk plane, which vanishes in an aligned rotator for which $\theta_0 =0.$
 Thus, the axis-ratio does not grow in model A00, for which the initial magnetic field lines are parallel to the rotation axis.
 This tendency is more marked in group C.
 We will discuss the correlation of the axis-ratio and the initial angle $\theta_0$ in \S 3.2.
 Although the high-density region rotates slightly around the rotation axis $\vect{\Omega}$ in the isothermal collapse phase [Figure~\ref{fig:1} (a)--(c)], it begins to rotate in the adiabatic phase [Figure~\ref{fig:1} (c) and (d)] when the gravitational collapse is suppressed by the thermal pressure.
 Comparing Figures~\ref{fig:1} (c) and (d), we can see the magnetic field lines and the disk-normal rotate around the $z$-axis.

\begin{figure}  
\epsscale{1.0}
\plotone{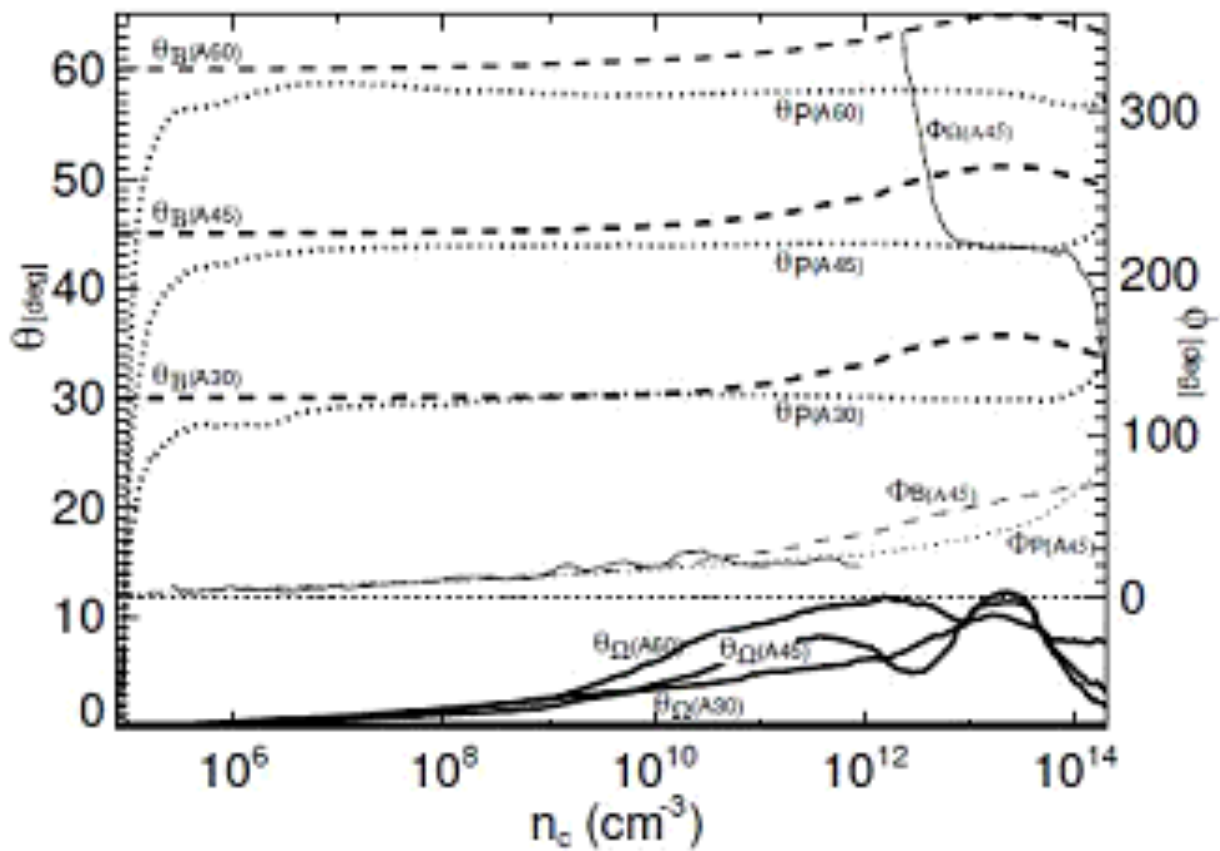}
\caption{
Angles $\theta_B$ (broken line), $\theta_\Omega$ (solid line), and $\theta_P$ (dotted line) for models A00, A30, A45, and A60, plotted against the central number density ($n_c$).
The angles $\phi_B$ (thin broken line), $\phi_\Omega$ (thin solid line), and $\phi_P$ (thin dotted line) for model A45 are also plotted.
The left and right ordinates indicate the angles $\theta$ and $\phi$, respectively.
The abscissa represents the central number density.
}
\label{fig:4}
\end{figure}

 Figure~\ref{fig:4} shows the evolution of the angles $\theta_B$, $\theta_\Omega$, and 
$\theta_P$ against the central density for models A00, A30, A45, and A60.
 The angles $\phi_B$, $\phi_\Omega$, and $\phi_P$ for model A45 are also plotted.
 In this figure, we cannot see any differences in the angles $\theta_B$, $\theta_\Omega$, and $\theta_P$ of model A00, because these angles are all zero. 
 The angles $\theta_B$ (broken lines) show that the magnetic field hardly changes with respect to the $z$-axis until the end of the calculation period.
 Thus, the magnetic field maintains its initial zenithal angle for group A.
 In contrast, the disk normal changes to the direction of the magnetic field after the central density exceeds $n \gtrsim 10^6 - 10^7 \cm$ for all models A00, A30, A45, and A60.
 Hence, we can see that the disk is formed perpendicular to the magnetic field, irrespective of the initial angle $\theta_0$.
 On the other hand, the rotation axis maintains its initial direction ($z$-axis) in the isothermal phase and then begins to move away from the $z$-axis in the adiabatic phase. 
 The angles $\phi_B$ and $\phi_P$ increase slightly in the isothermal phase because  the clouds rotate slowly.
 Figures~\ref{fig:2} and \ref{fig:4} clearly show that the evolutions of the oblateness and the angles $\theta_B$, $\theta_\Omega$, and $\theta_P$ do not qualitatively depend on the initial angle $\theta_0$, while the angles of the disk normal do depend on $\theta_0$.

   At the end of the calculation period for model A45, a disk structure is found in the region $r \lesssim 200$ AU, where the cloud has an oblateness of $\ob \ge 1.5$. 
   In this region, the direction of the disk normal varies depending on the disk scale.
   The disk normal is directed toward $\theta_P \sim 45\degr$, which is almost parallel to the magnetic field for $r \lesssim 50$ AU.
   On the other hand, for a disk in the range $r \simeq 100$ - $200$ AU, $\theta_P$ increases from $\sim 55\degr$ ($r\lesssim 100$ AU) to $\sim 60 \degr$ ($r\lesssim 200$ AU).
   Thus, the disk normal gradually becomes inclined and approaches the magnetic field direction, moving towards the center.  
   The magnetic field strength also depends on the scale.
   It increases as the center is approached and has a maximum at the center.

\begin{figure}  
\epsscale{1.0}
\plotone{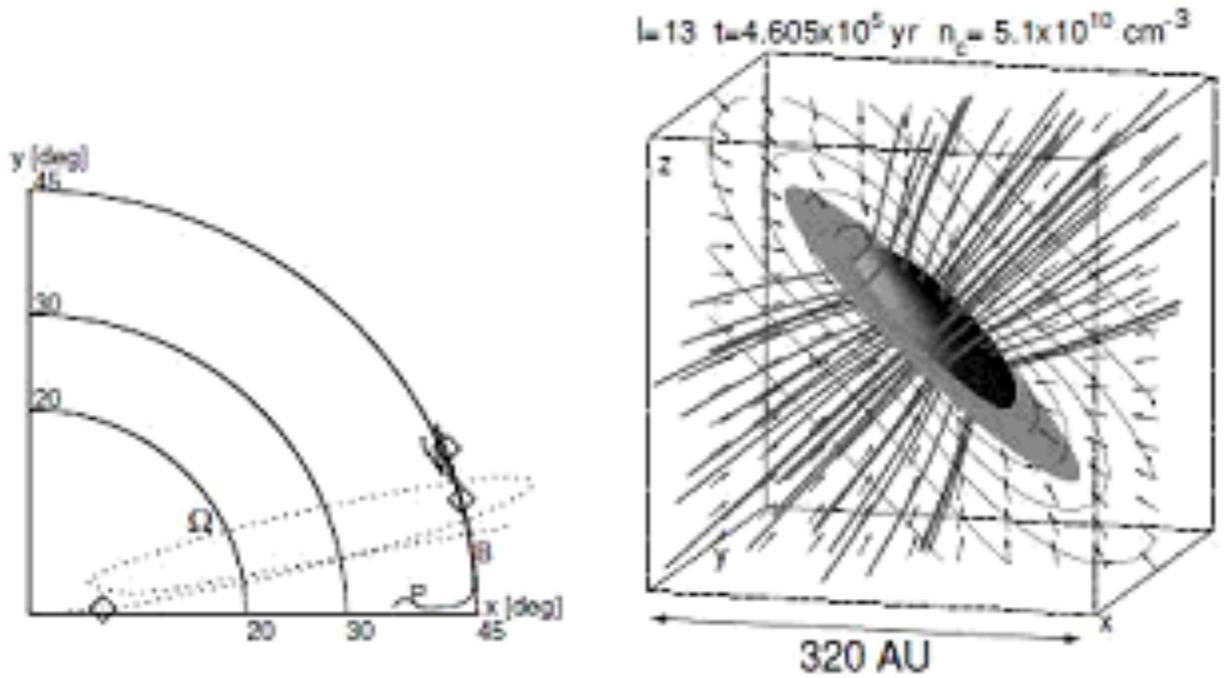}
\caption{
(Left) Loci of magnetic field, rotation axis, and disk normal for model B45.
The symbols $\diamond$ indicate the angles at the core formation epoch.
(Right) Structure of high density region ($n > 0.1 n_c$; isodensity surface), density contours (contour lines), velocity vectors (arrows), and magnetic field line (streamline) for $n_c = 5 \times 10^{10} \cm$.
}
\label{fig:5}
\end{figure}

   Next, we focus on group B ($\alpha=0.1$, $\omega = 0.01$) which has a magnetic field $10^{1/2}$-times stronger than group A.
The left panel of Figure~\ref{fig:5} shows the same information as Figure~\ref{fig:3} for model B45 ($\theta_0 = 45 \degr$).
 The cloud structure, magnetic field lines, and velocity vectors for model B45 at the end of the isothermal phase are shown in the right panel, for which the contour, streamline, isosurface, and other notations have the same meanings as in Figure~\ref{fig:1}.  
 In group B, a disk is formed perpendicular to the magnetic field, similar to group A.
 The direction of the magnetic field rotates around the $z$-axis, keeping the initial angle $\theta_0$, similar to model A45. 
 The directions of the magnetic field and the disk normal also coincide for model B45. 
 The angles $\phi_B$ and $\phi_P$, however, increase slightly in the isothermal phase, compared with model A45.
 Since the initial rotation speed $\omega = 0.01$ is common for groups A and B, this difference must be due to the growth rate of the angular velocity $\Omega$, which is smaller in group B than in group A, as shown in Paper II.
 The growth rates of the angular velocity ($\Omega$) and magnetic flux density ($B$) are large when the cloud collapses spherically ($\propto \rho^{2/3}$), while they are small when the cloud collapses laterally ($\propto \rho^{1/2}$) (Paper II). 
 A cloud with a strong magnetic field (group B) forms a disk earlier than that with a weak magnetic field (group A), and thus, the growth rate of the angular velocity in group B decreases from $\Omega \propto \rho^{2/3}$ to $\Omega \propto \rho^{1/2}$ at an earlier epoch than for group A. 
 Further, magnetic braking is more effective in group B than in group A, since the initial magnetic field is stronger in group B.
 Model B45 has  $\theta_\Omega = 7 \degr$ at the end of isothermal phase, while model A45 has  $\theta_\Omega = 2 \degr$ (Table~\ref{table:results}).
 That is, the rotation axis is more inclined with respect to the $z$-axis in model B45 than in model A45. 
 This inclination is caused by magnetic braking (MT04).
 MT04 show that magnetic braking works more effectively for the perpendicular component of angular momentum to the magnetic field than for the parallel component.
 The angular velocity in model B45 is slower than model A45 for the following two reasons, prompt disk formation and effective magnetic braking.
 The direction of the rotation axis oscillates violently around the $z$-axis in the adiabatic phase (Figure~\ref{fig:5}) because  magnetic braking is so effective, as noted in MT04 (model MF) and Paper II.
In conclusion, groups A and B exhibit a disk perpendicular to the magnetic field line irrespective of the initial angle $\theta_0$.

\subsubsection{Support-sufficient models}
\begin{figure}  
\epsscale{1.0}
\plotone{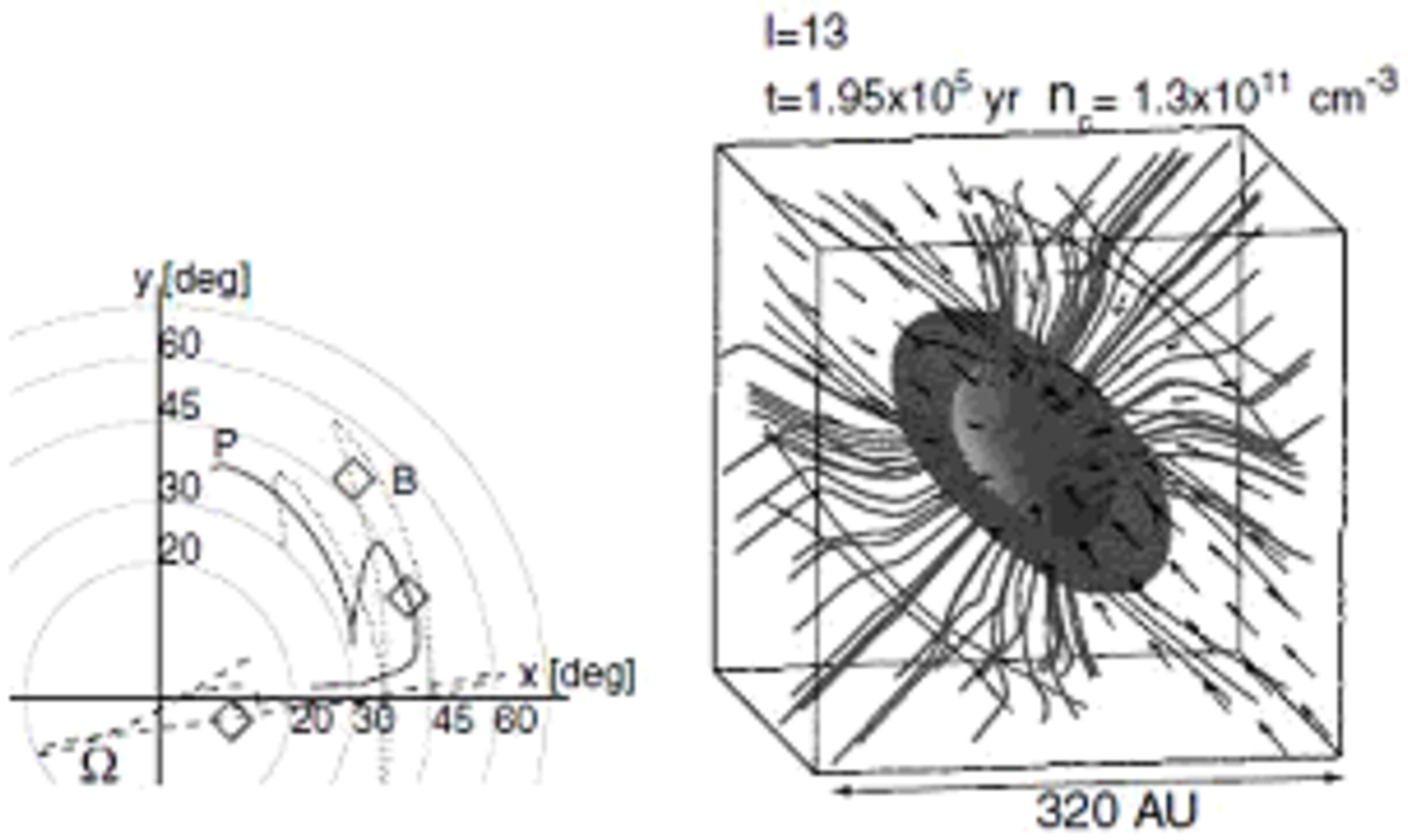}
\caption{
As Figure~5 for model F45
at $n_c = 1.3 \times 10^{11} \cm$ (right).
}
\label{fig:6n}
\end{figure}

 Group F models are magnetic-force dominant, similar to groups A and B but with a strong magnetic field.
 Group F has the parameters $\alpha = 1$ and $\omega = 0.05$.
 As group F clouds have a stronger magnetic field and rapid rotation at the initial stage, to promote cloud contraction we set a larger density enhancement factor for group F ($f = 5$) compared with that of groups A and B ($f= 1.68$).
 We can confirm that the density enhancement factor does not greatly affect cloud evolution (see models A45 and [n], or models B45 and [o] in Table~\ref{table:results}).
 The models in group F form disks through the magnetic force, as for models A45 and B45.
 For group F models, the cloud collapses along the magnetic field lines and contraction crossing the magnetic field lines is suppressed by the strong Lorentz force, while for models A45 and B45, the cloud collapses spherically. 
 The locus of the disk normal ($\vect{P}$) traces that of the magnetic field ($\vect{B}$) in the left panel of Figure~\ref{fig:6n}.
 The rotation axis is once inclined to $\theta_{\Omega} \simeq 70\degr$ in the isothermal phase, and then reverts  to $\theta_{\Omega} = 10\degr$ at the end of the isothermal phase.
 The loci of $\vect{B}$, $\vect{P}$, and $\vect{\Omega}$ are similar to those of models MF70 and MF80 of MT04, which have parameters ($\alpha$, $\omega$) = (0.76, 0.14).
 Although the three vectors ($\vect{B}$, $\vect{P}$, and $\vect{\Omega}$) do not completely converge to align in models MF70 and MF80, these vectors are roughly parallel, as seen in the left panel of Figure~\ref{fig:6n}.
 The right panel of Figure~\ref{fig:6n} shows the cloud structure, magnetic field lines, and velocity vectors at the end of the isothermal phase.
 This panel shows that a disk forms perpendicular to the magnetic field lines, as for groups A and B.

The axis-ratio grows to $\ar = 0.4$ in model F45 at the end of the isothermal phase (Table~\ref{table:results}).
The non-axisymmetric structure is caused by the centrifugal force whose rotation vector is perpendicular to the magnetic field in magnetic-force-dominant models. The axis-ratio begins to grow after a thin disk is formed. 
The axis-ratios in group F are larger than those of groups A and B (see Table~\ref{table:results})
as the disk formation epoch in group F is earlier than that of groups A and B.

In groups A, B, and F (magnetic-force-dominant models), a disk is formed perpendicular to the local magnetic field.
 Our results agree qualitatively with those of MT04 (models SF, MF, and WF).

\subsection{Rotation-Dominant Disks}
 In this subsection, we show the cloud evolution of groups C and E.
 These groups are expected to form rotation-dominant disks in the isothermal phase since $\Omega_0/B_0 > 0.39 G^{1/2} c_s^{-1}$.
 The cloud collapses along the rotation axis (vertical collapse) in group C (support-sufficient models; $ F \gtrsim 1$), while it collapses spherically in group E (support-deficient models; $F \lesssim 1$).

\subsubsection{Support-deficient models}
   The models of group E are rotation dominant ($\Omega_0/B_0 > 0.39 G^{1/2} c_s^{-1}$), although they have a slow rotation rate of $\omega = 0.05$.
  Compared with group A, group E has a magnetic field $10^{1/2}$-times smaller but a 5-times larger angular velocity.
   Model E45 has parameters $\alpha = 0.001$, $\omega=0.05$, and $\theta_0 = 45 \degr$.

\begin{figure}  
\plotone{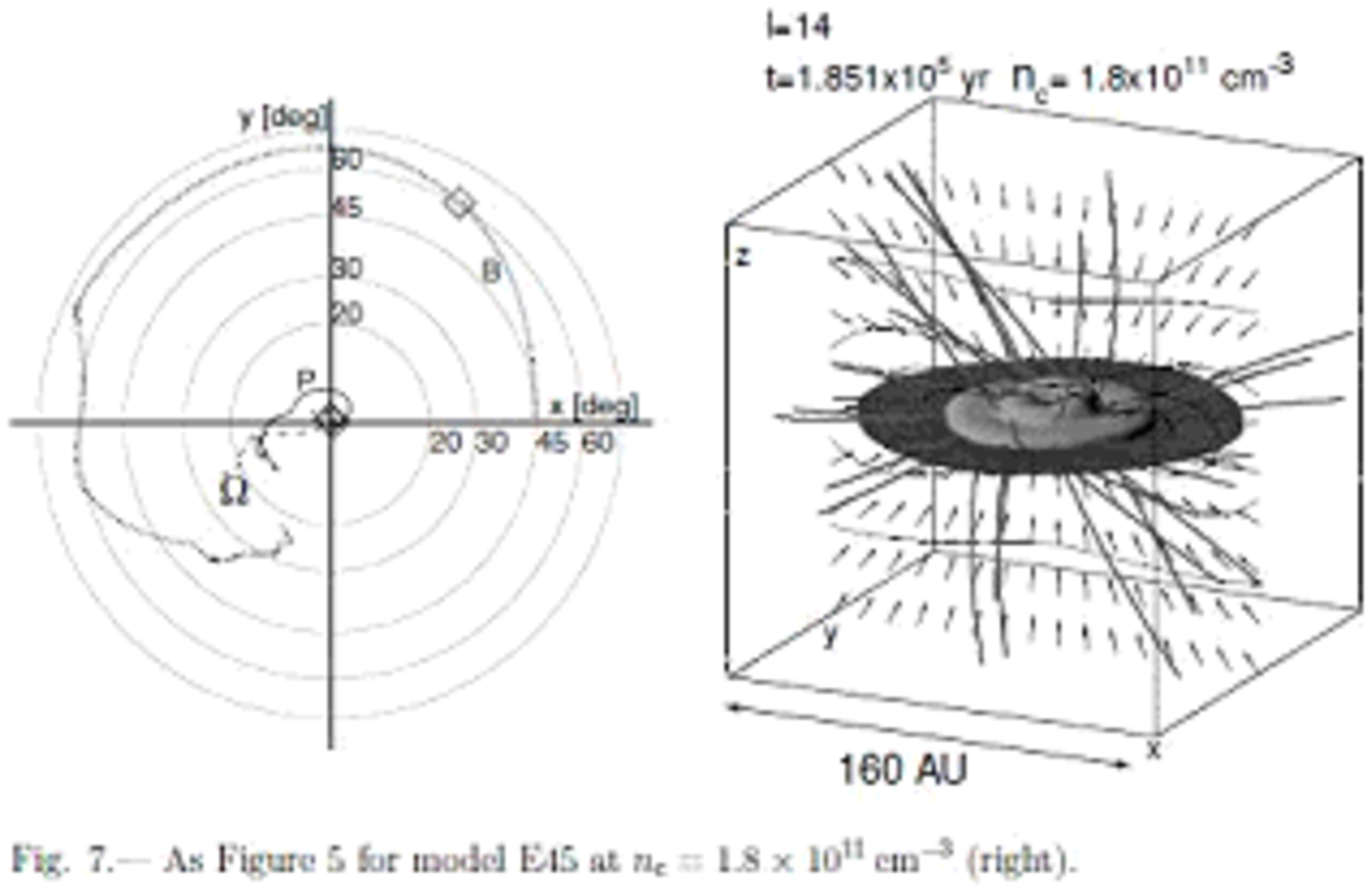}
\caption{
As Figure~5 for model E45 at $n_c = 1.8 \times 10^{11} \cm$ (right).
}
\label{fig:7n}
\end{figure}

   The models of group E form a disk through the centrifugal force.
  The left panel of Figure~\ref{fig:7n} shows that the rotation axis maintains its initial direction $\theta_\Omega \simeq 0\degr$ and the disk normal is also parallel to the $z$-axis (the rotation axis) in the isothermal phase.
   Therefore a disk forms perpendicular to the rotation axis, not along the magnetic field.
   The angle $\theta_B$ increases gradually in the isothermal phase.
   This means that the magnetic field tends to be aligned along the direction {\em perpendicular} to the rotation axis.
   However, the isothermal phase ends before the magnetic field lines are completely directed in the perpendicular direction.
   Since the models in group E have initial states inside the \bw relation line,\footnote{The cloud collapses slowly  in a spherically symmetric fashion inside the \bw relation line.  Otherwise, the cloud rapidly collapses along the vertical axis when the model is distributed outside the \bw relation line.}
 the magnetic field gradually becomes inclined as the cloud collapses slowly in a spherically symmetric fashion.
  Although the direction of the magnetic field does not coincide with the direction of rotation and of the disk normal in the isothermal phase, the magnetic field, rotation axis, and disk normal begin to converge with each other after the core formation epoch.
 As shown in MT04, magnetic braking works preferentially for the component of the angular momentum perpendicular to the magnetic field, which drives the alignment of \vect{B} and \vect{\Omega}.

\subsubsection{Support-sufficient models}
\begin{figure}
\plotone{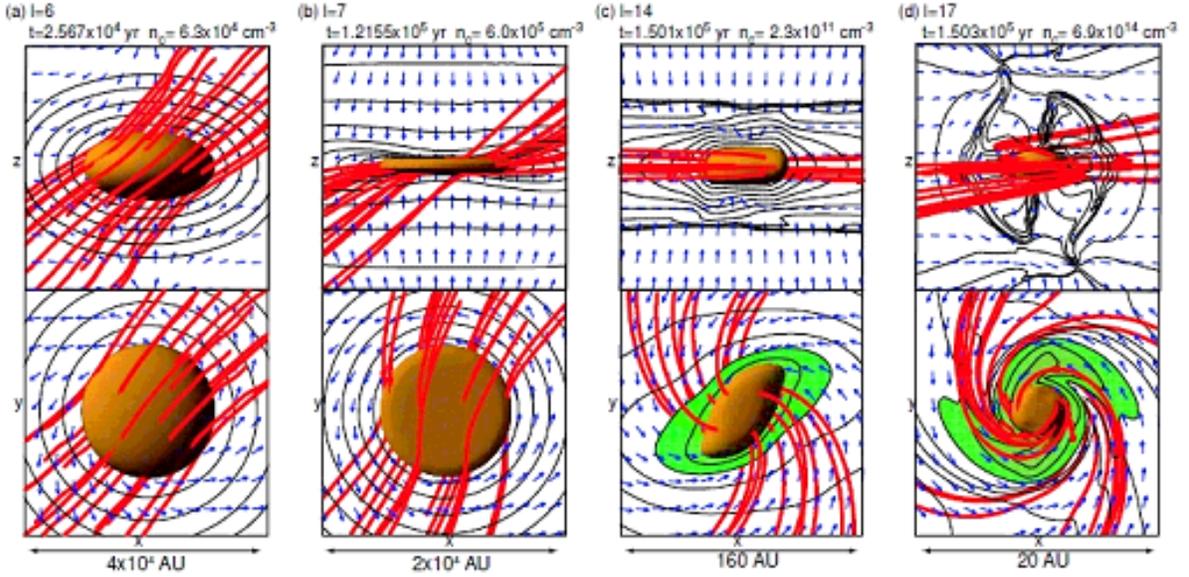}
\caption{
Snapshots of model C45.
The structure of the high-density region ($n > 0.1 n_c$\,; isodensity surface), density contours (contour lines), velocity vectors (arrows), and magnetic field lines (streamlines) are plotted in panels 
(a)--(d).
The upper panels show the view along the $y$-axis (edge-on view).
The lower panels show the view along the $z$-axis (face-on view).
 The green colored disk in panels (c) and (d) indicates the region of $n > (1/100)\, n_c$ on the mid-plane parallel to the disk-like structure. 
The panels show the stages (a) $n_c = 6.3 \times 10^4\cm$, (b) $n_c = 6 \times 10^5 \cm$, (c) $n_c = 2.3 \times 10^{11} \cm$, and (d) $n_c = 6.9 \times 10^{14} \cm$. 
The level of the finest grid ($l$), elapsed time from the beginning ($t$), and central number density ($n_c$) are listed at the top of each panel.
The size of the grid is also shown.
}
\label{fig:8n}
\end{figure}

 Figures~\ref{fig:8n} (a)--(d) show the cloud evolution of model C45 in views along the $y$-axis (edge-on view; upper panels) and along the $z$-axis (face-on view; lower panels).
 Model C45 has the parameters $\alpha=0.01$, $\omega=0.5$, and $\theta_0=45 \degr$.
 Compared with group A, group C has the same magnetic field but has a 50-times larger angular 
velocity.

\begin{figure}  
\epsscale{0.5}
\plotone{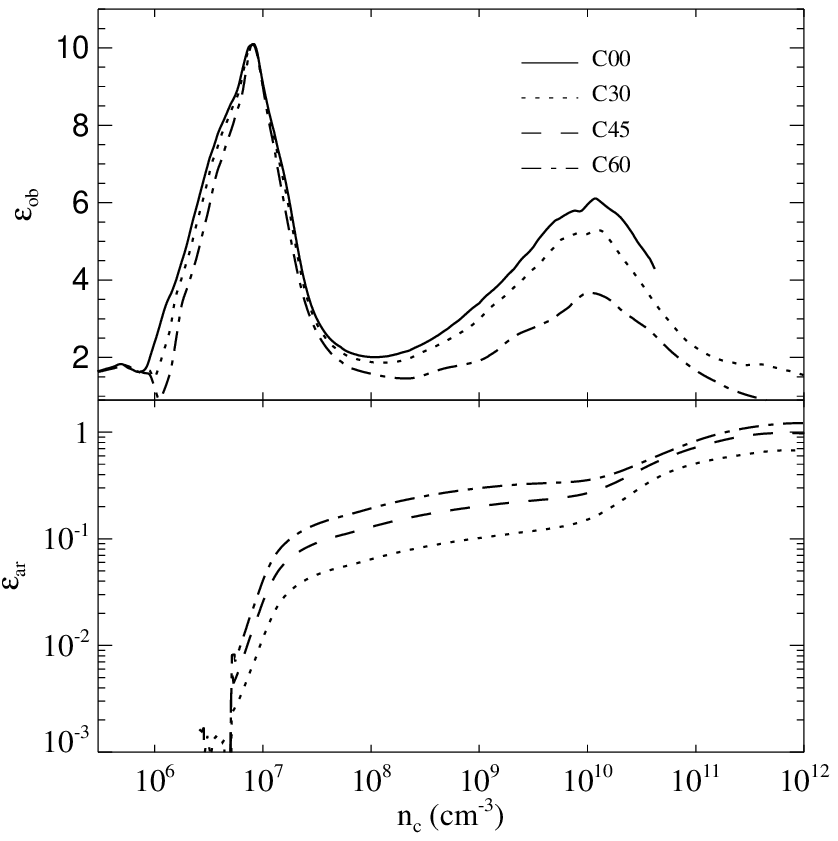}
\caption{
As Figure~2 for group C (models C00, C30, C45, and C60). 
}
\label{fig:9n}
\end{figure}

 Figure~\ref{fig:8n} (a) shows the cloud structure at $n_c = 6.3 \times 10^4 \cm$.
 The spherical cloud collapses along the rotation axis (i.e. the $z$-axis), and then an oblate core forms at the center (upper panel).
 The magnetic field lines are slightly squeezed at the center (upper panel) and are rotated $\phi_B \simeq 45 \degr$ from the initial stage $\phi_B=0\degr$ [Figure~\ref{fig:8n} (a) lower panel].
 Figure~\ref{fig:8n} (b) shows the core shape at $n_c = 6 \times 10^5 \cm$.
 A thin disk is formed in the $x$-$y$ plane.
 In this model, an extremely thin disk forms in the isothermal phase ($\ob \simeq 10 $ at $n_c = 9 \times 10^6 \cm$, seen in Figure~\ref{fig:9n}).
 In model CS of Paper II, a thin disk forms promptly in the early isothermal phase, because the lateral collapse is suppressed by a strong centrifugal force and hence the cloud collapses only vertically along the $z$-axis.
 In model C45, the cloud similarly collapses vertically and forms a disk promptly in the isothermal phase.
 The weak magnetic field in this model hardly affects the cloud evolution.
 Moreover, the magnetic field is compressed in the direction of the rotation axis and begins to run along the disk surface, as shown in Figure~\ref{fig:8n} (b).
 Outside the thin disk, the magnetic field lines run in the direction $\theta_B \simeq 45\degr$.
  That is, the magnetic field lines emerge from the lower-left side of the cloud and escape from the disk in the upper-right direction.
 This configuration of the magnetic field appears only in non-aligned rotators.
 In contrast, the disk is vertically threaded by the magnetic field along the rotation axis in model C00.
 This configuration of the magnetic field is seen in all the models studied in Paper II, which was restricted to aligned rotators.

\begin{figure}  
\plotone{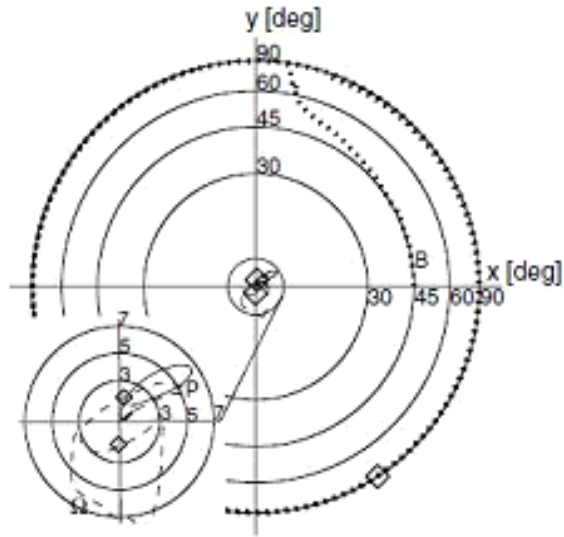}
\caption{
As Figure~5 (left) for model C45.
The lower left inset is an enlarged view of the center.
}
\label{fig:10n}
\end{figure}

 Figure~\ref{fig:10n} shows the direction of $\vect{B}$, $\vect{P}$, and $\vect{\Omega}$ for model C45.
 The inset at the lower-left corner is an enlarged view of the center. 
 This shows that the direction of the magnetic field gradually moves away from the $z$-axis and towards $\theta_B \simeq 90 \degr$.
   Then, the direction of the magnetic field rotates around the $z$-axis, keeping an angle of $\theta_B \simeq 90 \degr$. 
 On the other hand, the rotation axis hardly changes its direction from the initial state and remains directed along the $z$-axis. 
 The disk normal is also oriented along the $z$-axis (i.e. the rotation axis).
 From Figure~\ref{fig:8n} (b), it can be seen that a disk forms by the effect of the rotation and the disk normal direction coincides with the rotation axis.

 Figure~\ref{fig:8n} (c) shows the central region at the core formation epoch ($n_c = 2.3 \times 10^{11} \cm$).
 It can be seen from this figure that a non-axisymmetric structure has formed and the central core has changed its shape from a circular disk [lower panels of Figures~\ref{fig:8n} (a) and (b)] to a bar [Figure~\ref{fig:8n} (c) lower panel].
 The magnetic field lines run laterally, i.e. $|B_r|$, $|B_\phi| \gg |B_z|$, in the adiabatic phase [Figure~\ref{fig:8n} (c) and (d)].
 Figure~\ref{fig:8n} (d) shows an adiabatic core when the central density has reached $n_c = 6.9 \times 10^{14} \cm$.
   A spiral structure is seen in this figure, which indicates that a non-axisymmetric pattern has formed, even if no explicit non-axisymmetric perturbation is assumed  at the initial stage. 
 (Although the non-axisymmetric patterns also appear in some models of Papers I--III, it should be noted that these patterns are due to a non-axisymmetric perturbation added to the density and magnetic field at the initial stage.) 
 The magnetic field lines are considerably twisted in Figure~\ref{fig:8n} (d).
 It should be noted that in this model, the inclined magnetic field induces non-axisymmetric perturbations, on behalf of the initial explicit perturbation.

\begin{figure}  
\epsscale{0.5}
\plotone{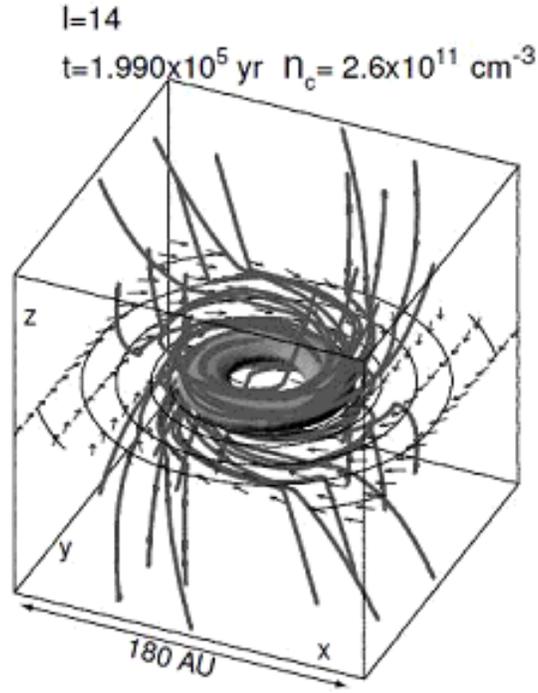}
\caption{
As Figure~1 for C00
at the core formation epoch.
}
\label{fig:11n}
\end{figure}

   Figure~\ref{fig:11n} shows the magnetic field lines, the shape of the core, and the velocity vectors on the $z=0$ plane in the adiabatic phase for model C00. 
   This figure shows that a ring is formed, as found in Paper III, without any growth of a non-axisymmetric pattern.
   In Papers I--III, we assumed a cylindrical cloud in hydrostatic equilibrium, in which the magnetic field and angular velocity are functions of the radius $r$ in cylindrical coordinates.
   On the other hand, the cloud is assumed to be spherical with a uniform magnetic field and angular velocity at the initial stage in model C00.
   In spite of these differences, a similar ring structure appears in both models C00 and CS of Paper II.
   The lower panel of Figure~\ref{fig:9n} plots the evolution of the axis-ratio against the central density for group C.
   The axis-ratios for models C30, C45, and C60 begin to grow after a thin disk is formed ($n_c \gtrsim 5 \times 10^6 \cm$) and reach $\ar \simeq 0.5$ at the core formation epoch.
   The axis-ratio grows to $\ar \simeq 1$ at $n_c = 10^{12} \cm$ in models C30, C45, and C60, while no non-axisymmetric pattern appears in model C00.
   This shows that the non-axisymmetric pattern arises from the anisotropy of the 
Lorentz force around the rotation axis.
   A bar structure is formed by the non-axisymmetric force exerted by the inclined magnetic field, as shown in Figures~\ref{fig:8n} (b)--(d). 
   This is  confirmed by the fact that the short axis of the density distribution 
on the $z=0$  plane (the disk mid-plane) and the bar pattern rotate together with the magnetic field lines.
  The axis-ratio (the non-axisymmetricity) grows in proportion to $\rho^{1/6}$ ($10^7 \lesssim n_c \lesssim 10^{10} \cm$ in the lower panel of Figure~\ref{fig:9n}), as found by \citet{hanawa99}.
  Since the lateral component of the magnetic field ($|\vect{B}|\, {\rm sin} \,\theta_0$)  is large (Figure~\ref{fig:9n} lower panel), the axis-ratio grows more in models with large $\theta_0$.

\begin{figure}  
\epsscale{0.8}
\plotone{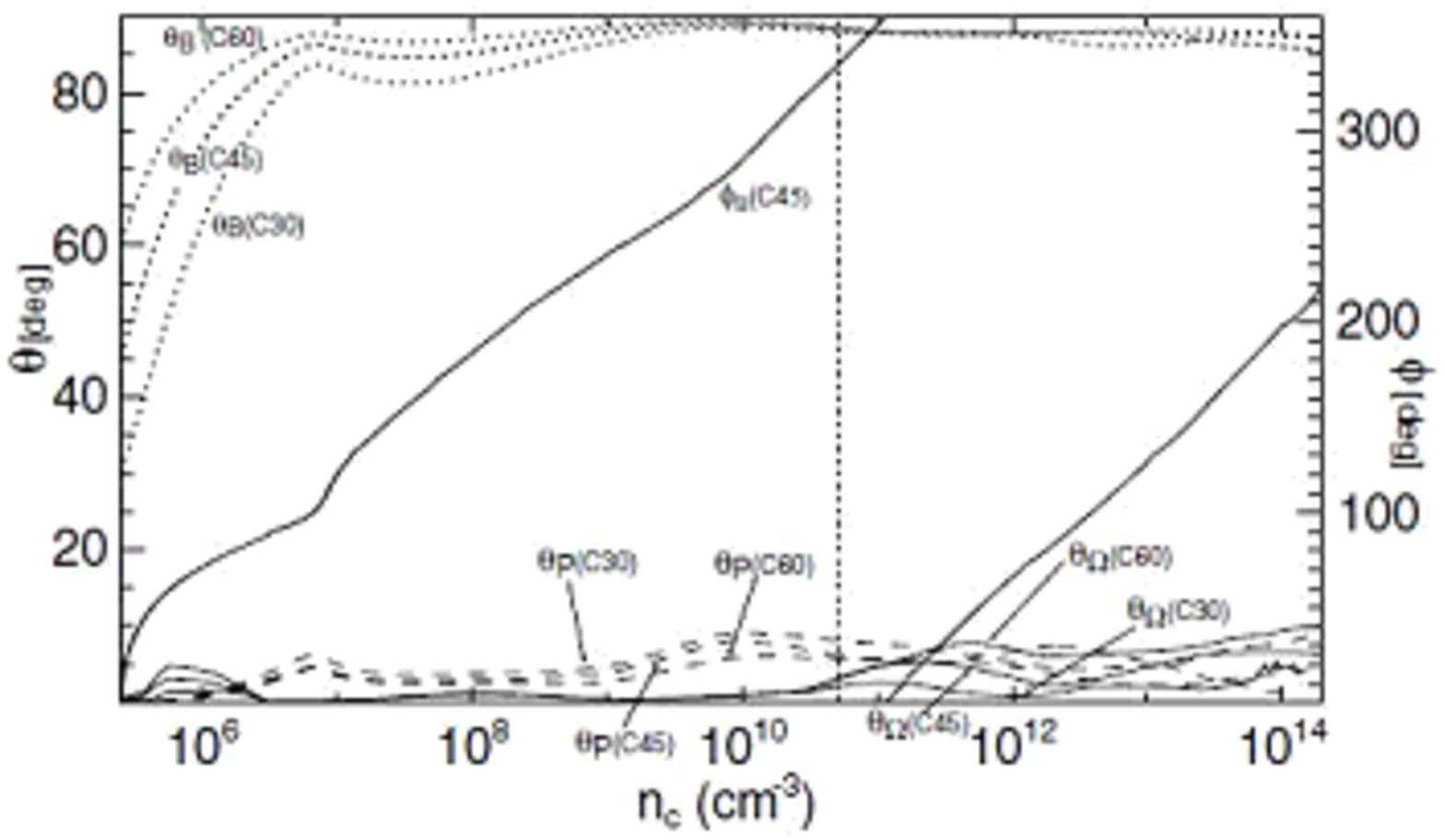}
\caption{
As Figure~4 for group C (models C00, C30, C45, and C60).
}
\label{fig:12n}
\end{figure}

   The evolution of the angles $\theta_B$, $\theta_\Omega$, $\theta_P$, and $\phi_B$  for group C are plotted against the central density in Figure~\ref{fig:12n}.
   The angle between the magnetic field and $z$-axis becomes $\theta_B \simeq 90\degr$ even in the early phase of isothermal collapse for all the models C30, C45, and C60.
   The rotation axis and the disk normal maintain their angles $\theta_\Omega, \ \theta_P \simeq 0\degr$. 
   Figures~\ref{fig:4} and \ref{fig:12n} show that in both magnetic- and rotation-dominant models the directions of the magnetic field, rotation axis, and disk normal are qualitatively the same for  models with the same $\alpha$ and $\omega$, irrespective of $\theta_0$ in the range $30\degr \le \theta_0 \le 60\degr$.

   Figure~\ref{fig:13n} shows the magnetic field lines, velocity vectors, and density distribution for the epoch $t=1.52\times 10^6$ yr ($n_c = 1.5 \times 10^9 \cm$) for model C30.
   Note that the box scale and level of grid are different for each panel.
   The spatial scale of each successive panel is different by a factor of four 
   and thus the scale between panels (a) and (d) is different by a factor of 64.
   The magnetic field has an angle $\theta_B' \sim 30\degr$ in panel (a), where  $\theta_B'$ is defined as the angle between the volume average magnetic field in the grid and the $z$-axis. 
   Although the magnetic field lines in the thin disk run parallel to the disk surface ($\theta_B \simeq 90\degr$; Figure~\ref{fig:13n} {\it b}),  the magnetic field lines outside the disk preserve the ambient direction of $\theta_B'  \simeq 30\degr$.
   Closer to the cloud center, the magnetic field lines are twisted near the disk surface in the azimuthal direction, as seen in Figures~\ref{fig:13n} (c) and (d).
   As a result, the directions of the magnetic field are considerably different for different scales or densities even in the same cloud.

\subsection{Disk Formation Affected Both by  Magnetic Field and Rotation}
   In this subsection, we detail the evolution of group D, in which both the magnetic field and angular velocity are crucial for cloud evolution and disk formation.
   Thus, group D is located near the border between the magnetic-force- and rotation-dominant models, $\Omega_0/B_0 \simeq 0.39 G^{1/2} c_s^{-1}$, and has the parameters $\alpha = 1$ and $\omega=0.5$.
   Group D has the same angular velocity as group C, but has a 10-times larger magnetic field. 
   The clouds in this group have both a strong magnetic field and rapid rotation.

   The left panel of Figure~\ref{fig:14n} shows the directions of $\vect{B}$, $\vect{\Omega}$, and $\vect{P}$ for model D45.
   It can be seen that the direction of the magnetic field oscillates in the range   $\theta_B \simeq 45 \degr$ -- $70 \degr$.
   Although the direction of the disk normal approaches the magnetic field direction,  they are not completely aligned, as is the case for groups A and B.  
   The direction of the rotation axis also oscillates considerably in the isothermal phase and is at $\theta_\Omega = 63 \degr$ at the core formation epoch.
   This value is almost the same as the angles of the magnetic field ($\theta_B = 51 \degr$) and the disk normal ($\theta_P = 49 \degr$), but the azimuthal coordinate  $\phi_\Omega =39 \degr$ is very different from those of the magnetic field ($\phi_B = 283\degr$) and the disk normal ($\phi_P = 275\degr$).
   Thus, the direction of the rotation axis differs greatly from those of the magnetic field  and disk normal.
   The direction of the disk normal is nearer to the magnetic field ($\psi_{BP} = 6\degr$) than the rotation axis
($\psi_{\Omega P} = 85\degr$) at the core formation epoch, as shown in Table~\ref{table:results}.
   Thus, the disk normal seems to be parallel to the magnetic field for model D45 (Figure~\ref{fig:14n} right panel). 
   On the other hand, the disk normal is nearer to the rotation axis ($\psi_{\Omega P} = 6 \degr$) than the magnetic field ($\psi_{B P} = 72 \degr$) in model D30.
   In model D60, the disk normal is close to both the magnetic field ($\psi_{B P} = 29 \degr$) and the rotation axis ($\psi_{\Omega P} = 33 \degr$).
   In these clouds, the direction of the magnetic field, rotation axis, and disk normal oscillate in the isothermal phase.
   Whether a disk is aligned perpendicularly to the magnetic field or rotation motion depends on $\theta_0$.
   The cloud evolutions are influenced both by the magnetic field and the centrifugal force in group D.
   Therefore, we cannot clearly classify models in group D into either magnetic-force- or rotation-dominant models.

\section{Magnetic Flux--Spin Relation}
\subsection{Amplification of the Magnetic Field and Angular Velocity}
   The magnetic field strength and angular rotation speed increase as a cloud collapses.
   We have found in Paper II that the magnetic field strength normalized by the gas pressure and the angular velocity normalized by the free-fall timescale satisfy Equation~(\ref{eq:UL}) after a contracting disk forms in the isothermal phase for an aligned rotator model.
   In this subsection, we investigate the above relation for the case when the magnetic field is not necessarily parallel to the rotation axis (non-aligned rotator models).
   The evolution loci of the cores are plotted in Figure~\ref{fig:15n}, where the horizontal and vertical axes were calculated using the central values of $\rhoc$, $B_c$, and $\Omega_c$.
   In Figure~\ref{fig:15n} a thick band indicating the equation 
\begin{equation}
   \frac{\Omega_c^2}{(0.2)^2 \; 4 \pi G \rho_c} +
   \frac{B_{\rm c}^2}{(0.36)^2 \; 8 \pi c_s^2 \rho_c} =1
   \label{eq:UL2}
\end{equation}
   is also drawn, where the numerators of the left-hand side are defined as $B_c = (B_{x,c}^2 + B_{y,c}^2 + B_{z,c}^2)^{1/2}$ and $\Omega_c = (\Omega_{x,c}^2 + \Omega_{y,c}^2 + \Omega_{z,c}^2)^{1/2}$, where the suffix $c$ indicates the values at the center.

   First, we consider the aligned rotator models (models with $\theta_0 = 0$) and compare them with those of Paper II.
   Comparing the solid lines ($\theta_0 = 0 \degr$ models) of Figure~\ref{fig:15n}  with  those of Figure~12 in Paper II, we can see the evolution locus is almost the same.
   That is, (1) the points move from the lower-left to the upper-right inside the \bw relation line, whereas (2) those distributed outside the line move from the upper-right to the lower-left; (3) the slope of each evolution path is $ d\log \Omega _{\rm c} / d \log B_{\rm c} \simeq 1 $ and (4) the evolution paths finally converge to Equation~(\ref{eq:UL2}) for the isothermal phase.
   However, the evolution paths for models C00 and D00, which are located outside the \bw relation line have slightly smaller angular velocities [$\Omega_c \, (4 \pi G \rhoc)^{-1/2} \simeq 0.1$--$0.15 $] than those of models C and D of Paper II [$\Omega_c \, (4 \pi G \rhoc)^{-1/2} \simeq 0.2$].
   This seems to be due to the fact that the initial cloud considered in this paper is more unstable against gravity than that of Paper II; the ratio of thermal energy to gravitational energy is $\alpha_0 = 0.168$ in the present paper, while $\alpha_0 \simeq 0.6$-$0.7$ in Paper II.
   Although the clouds collapse vertically (vertical collapse) for the models   C, D, C00, and D00 until the magnetic field and rotation satisfy the \bw relation, the vertical collapse overshoots the \bw relation line for initially unstable clouds in models C00 and D00.
   This difference can also be seen in comparing the density increase rate.
   The increase of the central density can be approximated by $\rho_c \simeq 5.1 / [4 \pi G (t-t_f)^2]$ in model C00, where $t_f$ is the time at which the central density becomes infinite for the isothermal phase.
 This density increase rate is 1.75-times slower than that of the similarity solution ($\rho_c = 1.667/[4 \pi G (t-t_f)^2]$; Larson 1969; Penston 1969).
On the other hand, the density increase can be approximated by $\rho_c \simeq 6.2 / [4 \pi G (t-t_f)^2]$ for model C of Paper II, giving a density increase rate $(6.2/5.1)^{1/2} \simeq 1.1 $-times faster in model C00 than in model C.
   This is a natural consequence of the lower $\alpha_0$ of model C00.
   This difference is also seen in the evolution of the oblateness.
   For example, the oblateness reaches $\ob \simeq 10$ in model C00, while it reaches only $\ob \simeq 4$ in model C.
   Thus, a thinner disk is formed in model C00, which has a more unstable initial state.

   The evolution locus of model F00 moves towards the upper-right in the period $n_c \lesssim 10^6 \cm$ in Figure~\ref{fig:15n}.
   This indicates that the cloud collapses spherically in this period, as for models A00 and B00, because the cloud is more unstable against the gravity than those of groups C and D.
   Group F has the same thermal energy as groups C and D, but has smaller rotational and magnetic energies, as shown in Table~\ref{table:init}.
   Thus, in model F00, the cloud collapses spherically in the early phase of the isothermal collapse.
   However, the collapse becomes anisotropic in the late phase of isothermal collapse because the magnetic force becomes effective.
   Although there are a few differences between the models, the convergence to the \bw relation curve is evident, irrespective of the initial cloud shape and the distributions of the density, magnetic field, and angular velocity when the magnetic field is parallel to the rotation axis.
  This is natural because the \bw relation is satisfied for the central part of the cloud and information on the outer part of the cloud is lost as the cloud collapses in the isothermal phase, as noted by \citet{larson69}.

   Next, we consider the non-aligned rotator models ($\theta_0 \ne 0$).
   In Figure~\ref{fig:15n}, the dotted, broken, and dash-dotted lines show the evolution paths of models with $\theta$ = 30$\degr$, 45$\degr$, and 60$\degr$, respectively.
   The points located inside the \bw relation line (non-aligned rotator models in groups A, B, and E) move from the lower-left to the upper-right, as for the aligned rotator models.
   These models have almost the same loci as the aligned rotator models and converge to the \bw relation line.
   On the other hand, models located outside the \bw relation line (non-aligned rotator models in groups C, D, and F) show different evolutions than the aligned rotators.
   The non-aligned rotator models C30, C45, and C60 evolve towards the {\em lower-right}, then reverse their direction after they reach the \bw relation line. 
   Thus, the evolution of the angular velocity in the non-aligned rotator models C30, C45, and C60 is the same as that of the aligned rotator model C00, while the evolution of the magnetic field is different.
   The magnetic field strength normalized by the gas pressure increases in non-aligned rotator models, but decreases in aligned rotator models.
   The evolution paths of non-aligned rotator models of group F (F40, F45, and F60) are toward the upper-left.
   Thus, the magnetic field strength normalized by the thermal pressure approaches the \bw relation line in these models, while the angular velocity normalized by the free-fall timescale continues to increase in the isothermal collapse phase. 
   The evolution paths for models D30, D45, and D60 oscillate in the $B_c\, (8\pi c_s^2 \rhoc)^{-1/2}$--$\Omega_c \, (4 \pi G \rhoc)^{-1/2}$ plane, moving away from the \bw relation line.

\subsection{Generalized Magnetic Flux--Spin Relation}
  The growths of the magnetic field strength and angular velocity depend on the geometry of the collapse (vertical collapse to form a disk, spherical collapse, or lateral collapse in a disk).
   Figure~\ref{fig:16n} is similar to Figure~\ref{fig:15n}, but with the magnetic field strength and the angular velocity parallel to the disk normal.
   That is,  $B_{cp}$ in the abscissa and $\Omega_{cp}$ in the ordinate are defined as
\begin{equation}
B_{cp} = \vect{B} \cdot \vect{p},
\label{eq:bp}
\end{equation}
\begin{equation}
\Omega_{cp} = \vect{\Omega} \cdot \vect{p},
\label{eq:wp}
\end{equation}
where $\vect{B}$, $\vect{\Omega}$, and $\vect{p}$ represent the magnetic flux density vector, angular velocity vector, and the unit vector of the disk normal.
   The starting points of each locus are different even for models with the same $\alpha$ and $\omega$ but different $\theta_0$ (c.f. A00 and A45), since the angle of the magnetic field at the initial stage and the formation epoch of the disk structure are dependent on $\theta_0$.
 We plotted the loci according to Equations~(\ref{eq:bp}) and (\ref{eq:wp}) subsequent to the formation of a disk structure.
  The figure shows that all the evolution paths of models with the same $\alpha$ and $\omega$ (e.g. A00 and A45) move in the same direction irrespective of $\theta_0$, even though the starting points are different.

   Clouds for the models inside the \bw relation line (support-deficient models: groups A, B, and E) evolve towards the upper-right, regardless of the initial angle $\theta_0$  in Figure~\ref{fig:16n}.
   Clouds having parameters inside the \bw relation line collapse spherically until the magnetic field strength and angular velocity reach the \bw relation line, as shown in \S 4.1.
   The clouds evolve isotropically (spherically), because the anisotropy caused by the magnetic or centrifugal force is not induced before the clouds reach the \bw relation line for small initial magnetic and rotational energies.
  Thus, the evolution direction of the non-aligned rotator models A45, B45, and E45 is the same as that of the non-aligned rotator models A00, B00, and E00 in Figure~\ref{fig:15n}, though the angles between the magnetic field and rotation axis are different.
   This is because the anisotropy grows only slightly during spherical collapse.

   For the support-sufficient models, the evolution paths of the non-aligned rotator models C45, D45, and F45 are not the same as the aligned rotator models C00, D00, and F00 in Figure~\ref{fig:15n}.
   The cloud collapses vertically (vertical collapse) for groups C, D, and F, as shown in \S 3.2 and Paper II.
   In the case of the evolution of a weakly magnetized cloud rotating rapidly, in which the magnetic field is not parallel to the rotation axis and the magnetic field does not affect the cloud evolution, as seen in model C45,
   the cloud collapses along the rotation axis and the lateral collapse is suppressed by the centrifugal force.
   The cloud then forms a disk perpendicular to the rotation axis.
   The magnetic field and angular velocity parallel to the disk normal increase slightly for this collapse ($\Omega_{cp}$ and $B_{cp} \approx$ constant).
   Thus, as the collapse proceeds, the evolution path is toward the lower-left, as shown in Figure~\ref{fig:16n} and also seen for model C00.   
  The magnetic field perpendicular to the disk normal (parallel to the disk) is amplified with the cloud collapse, when the cloud has a magnetic field that is not parallel to the disk normal at the initial stage.
   Including this component of the magnetic field, the evolution path is toward the lower-right, as shown in Figure~\ref{fig:15n} [the numerator of the second term of Equation~(12) increases].
   This is the reason why the normalized magnetic field strength in model C45 of Figure~\ref{fig:15n} increases in the isothermal collapse phase.
   The case for group F is similar to that of group C,
   however the cloud evolution is mainly controlled by the magnetic field for group F, not by the centrifugal force as in group C.
   Thus, the roles of the magnetic field and angular velocity are reversed.
   In group F clouds collapse along the magnetic field lines and disks are formed perpendicular to the  magnetic field.
   The angular velocity {\em perpendicular} to the magnetic field or disk normal is then amplified and the angular velocity parallel to the magnetic field increases slightly.

   In Figure~\ref{fig:16n}, the ordinate $\Omega_{cp} \, (4 \pi G \rhoc)^{-1/2}$ indicates the ratio of the rotation and gravitational energies, while the abscissa $B_{cp}\, (8\pi c_s^2 \rhoc)^{-1/2}$ indicates the ratio of the magnetic and thermal energies.
   Since these ratios decrease in the aligned rotator models in proportion to $\rhoc^{-1/2}$ for the vertical collapse phase, the thermal and gravitational energies catch up with the magnetic and rotational energies eventually.
   The cloud then reaches the \bw relation line and the geometry of the collapse changes from vertical to lateral in the disk.
   Thus, a balance between the magnetic, rotational, thermal, and gravitational forces is achieved in a collapsing cloud. 
   This type of evolution occurs in groups C, D, and F.

   When the evolution loci for groups A and B approach the \bw relation line, the evolution depends on $\theta_0$ even for the same $\alpha$ and $\omega$.
   For groups A and B, both ratios $\Omega_c \, (4 \pi G \rhoc)^{-1/2}$ and $B_c\, (8\pi c_s^2 \rhoc)^{-1/2}$ increase in proportion to $\rhoc^{1/6}$ for a spherical collapse in which the cloud has small magnetic and rotational energies.
   Thus, the magnetic and rotational energies become comparable to the gravitational and thermal energies during the contraction.
   The cloud then reaches the \bw relation line and the geometry of the collapse changes from spherical to lateral.
   As a result, the anisotropy in $\theta$ appears as the cloud reaches the \bw relation line.   
   Groups A and B form magnetic-force-dominant disks, in which the direction of the disk normal is controlled by the magnetic field.
   The magnetic field strength normalized by gas pressure does not increase or decrease after a cloud reaches the \bw relation line in Figure~\ref{fig:15n}.
   The angular velocity, however, can increase even after the cloud has reached the \bw relation line in Figure~\ref{fig:15n}, because the rotation axis is not parallel to the disk normal.
    Differences in the position of the end point in models A00, A30, A45, and A60 are caused by this  mechanism.
   Magnetic braking is also effective in these models.
   For these reasons, the evolution paths begin to diverge as they approach the \bw relation line.

   We have plotted the normalized magnetic field strength and angular velocity at the initial stage for models WF, MF, and SF of MT04 as diamonds in Figure~\ref{fig:16n}.
 MT04 shows that disks are formed perpendicular to the local magnetic field in all non-aligned rotator models of MT04.
   This is natural because the models WF, MF, and SF are distributed in the magnetic-force-dominant region in Figure~\ref{fig:16n}.

   In summary, the geometry of the collapse determines the amplification of the magnetic field and angular velocity.
   The gas cloud in the support-deficient region amplifies the magnetic field strength and angular velocity during the contraction, regardless of the initial angle $\theta_0$.
   In these models, aligned and non-aligned rotators evolve similarly.
   On the other hand, for the support-sufficient models, aligned and non-aligned rotators evolve differently.
   In the rotation-dominated models, the magnetic field perpendicular to the rotation axis is amplified.
   This plays a role as a non-axisymmetric perturbation in forming a bar or spiral structure.
   Even in non-aligned rotator models, the generalized magnetic flux--spin relation holds in contracting disks formed in the isothermal regime.

\subsection{Disk Formation by Magnetic Field or Rotation}
  We have shown that a cloud forms either a magnetic-force-dominant or a rotation-dominant disk according to its initial conditions and that cloud evolution can be well understood and classified using the generalized \bw relation.
  In this subsection, we show how we can specify the parameter regions where a disk is formed under the influence of either the magnetic field or under the influence of rotation.
   The magnetic field,  rotation axis, and  disk normal have different directions when the magnetic field is not parallel to the rotation axis at the initial stage, although they have are identical for an aligned rotator.
    We can assess the dominant force for disk formation using the evolution loci of the direction of the magnetic field, rotation axis, and disk normal.
  When a disk is formed perpendicular to the magnetic field, the disk normal moves in association with a small-scale magnetic field in magnetic-force-dominant models (Figure~\ref{fig:3}).  
  A similar evolution is seen in rotation-dominated models.  
   We compare the evolution of 16 models with different $\alpha$ and $\omega$ but the same $\theta_0$.
   We choose the initial angle between the magnetic field and rotation axis as $\theta_0 = 45\degr$, because the cloud evolution does not depend on $\theta_0$, as shown in \S 3.1 and \S 3.2.
   The angles ($\theta_B$, $\theta_\Omega$, $\theta_P$), ($\phi_B$, $\phi_\Omega$, $\phi_P$), and ($\psi_{B \Omega}$, $\psi_{B P}$, $\psi_{\Omega P}$); the dominant force for forming a disk ($B$ or $\Omega$); and the axis-ratio ($\ar$) at the end of the isothermal phase are all listed in Table~\ref{table:results}.
   The dominant force for disk formation is determined by the loci of the magnetic field, rotation axis, and disk normal for the isothermal phase.
    For example, the locus of the disk normal moves together with that of the magnetic field in magnetic-force-dominant models.

 The shapes of the clouds at the core formation epoch are shown in Figure~\ref{fig:17n}.
   In this figure, each panel is positioned based on the initial magnetic field strength and angular velocity.
   The figure shows that the disk normals are almost parallel to the $z$-axis in the upper-left region, while they are parallel to the magnetic field in the lower-right region.
   This is natural since the cloud is initially rotating rapidly and magnetized weakly in the upper-left region, while it rotates slowly and is magnetized strongly in the lower-right region.

   In order to compare the cloud evolutions with different initial angular velocities we focus on four models with $\alpha$ = 0.01 and different $\omega$  = 0.3, 0.1, 0.03, and 0.01 [models (b), (f), (j), and (n)], shown in Figure~\ref{fig:17n}. 
   These models are aligned in the second column of Figure~\ref{fig:17n}. 
   The disk normal is oriented with the $z$-direction, and the magnetic field lines are inclined from the $z$-axis in the models with large $\omega$ [(b) and (f)].
   Each model has a small angle between the rotation axis and the $z$-axis [$\theta_\Omega = 2 \degr$ (b),  $3 \degr$ (f),  2$\degr$ (j),  and 5$\degr$ (n) ].
 Thus, it is shown that the cloud evolves maintaining the direction of the initial rotation axis.
   On the other hand, the angle between the magnetic field and the $z$-axis increases with increasing initial angular velocity $\omega$ [$\theta_B$ =  $ 86 \degr$ (b), $83 \degr$ (f), 54$\degr$ (j), and 46$\degr$ (n)].
  The magnetic field in models (b) and (f) is almost perpendicular to the rotation axis and the disk normal ($\psi_{B \Omega}$, $\psi_{B P} \simeq 90 \degr$). 
  It appears that the disk is formed by the effect of rotation in models (b) and (f), because the angle between the rotation axis and disk normal is small [$\psi_{\Omega P} = 2\degr$ (b) and $6\degr$ (f)].
  On the other hand, the disk seems to be formed by the magnetic force in models (j) and (n), because the angles $\psi_{B P}$ [$\psi_{B P} = 25\degr$ (j) and $4\degr$ (n)] are smaller than those of $\psi_{\Omega P}$ [$\psi_{\Omega P} = 28\degr$ (j) and $39\degr$ (n)].
  The axis-ratio increases with increasing $\omega$ [$\ar = 0.51$ (b), 0.18 (f), $0.91\times 10^{-2}$ (j), and $6\times 10^{-3}$ (n)], because the disk forms  earlier in a model with larger $\omega$.

   Next, we focus on four models with the same $\omega = 0.1$ but different $\alpha$ = 0.001, 0.01, 0.1, and 1 [models (e), (f), (g), and (h)], in order to compare cloud evolution with different initial magnetic field strengths.
   These models are aligned in the second row of Figure~\ref{fig:17n}.
   The disk normals are considerably inclined from the $z$-axis in the models with strong magnetic field, (g) and (h).   
   This inclination indicates that the disk is formed by the effect of the magnetic force.
   The disk is perpendicular to the rotation axis in model (e) [($\psi_{BP}$, $\psi_{\Omega P}$) = (83$\degr$, 1$\degr$)], while the disk is perpendicular to the local magnetic field rather than the rotation axis in model (h) [($\psi_{BP}$, $\psi_{\Omega P}$) = (2$\degr$, 11$\degr$)].

  The angle between the rotation axis and the disk normal ($\psi_{\Omega P}$) is smaller in model (h) ($\psi_{\Omega P} = 11 \degr$) than in model (g) ($\psi_{\Omega P} = 48 \degr$).
  This seems to be due to the fact that the angular momentum perpendicular to the magnetic field in models with $\alpha > 0.1$ is effectively removed by the magnetic braking process, and thus the direction of rotation tends to incline to the magnetic field and the disk normal.
   As a result, the rotation axis is considerably inclined from the initial direction in models with a strong magnetic field 
[$\theta_{\Omega}$ = 0 (e), 3 (f), 16 (g), and 32 (h)].
   The magnetic field is parallel to the disk in models with small $\alpha$ [e.g. $\psi_{BP} = 83$ (e)], while the magnetic field maintains its initial direction in models with large $\alpha$ [e.g. $\theta_B = 43$ (h)].
   The azimuthal directions of the magnetic field ($\phi_B$) and disk normal ($\phi_P$) coincide in models (e), (f), (g), and (h).
   In these models, the gas contracts along the magnetic field line onto the disk mid-plane, then a non-axisymmetric structure (i.e. bar structure) is formed perpendicular to the magnetic field, as discussed in \S 3.2.
   The non-axisymmetry tends to increase with the initial magnetic field strength [see models (e), (f), and (g)], except for model (h).

   The role of the magnetic field in magnetic-force-dominant models is similar to that of the centrifugal force in rotation-dominant models.
   The disk orientation is essentially determined by the direction of the dominant force.
  However, there is at least one quantitatively different point between the two types of models. 
  Namely, the angular momentum is transferred by the magnetic braking in the magnetic-dominant models.
   The dominant force ($B$ or $\Omega$) for disk formation is summarized in Table~\ref{table:results} and Figure~\ref{fig:17n}.
   The shadowed region in the lower-right part of Figure~\ref{fig:17n} indicates disks formed by the Lorentz force, while the upper-left region indicates disks formed by the centrifugal force.
   A broken line between these two indicates the border between the magnetic-force-dominant and rotation-dominant disks, which is well fitted by
   \begin{eqnarray}
      \frac{\Omega_0 }{B_0 }   \simeq 
      0.39 \, G^{1/2}{ c _{\rm s}}^{-1},
      \label{eq:UL3}
   \end{eqnarray}
   similar to the result for aligned rotators.

   Cloud evolution can be classified into four patterns using the generalized \bw relation curve 
\begin{equation}
   \frac{\Omega_{cp}^2}{(0.2)^2 \; 4 \pi G \rho_c} +
   \frac{B_{cp}^2}{(0.36)^2 \; 8 \pi c_s^2 \rho_c} =1,
   \label{eq:UL3b}
\end{equation}
and Equation~(\ref{eq:UL3}): (i) support-deficient, rotation-dominant models [inside the \bw relation line and above Equation~(\ref{eq:UL3})], (ii) support-deficient, magnetic-force-dominant models [inside the \bw relation line and below Equation~(\ref{eq:UL3})], (iii) support-sufficient, rotation-dominant models [outside the \bw relation line and above Equation~(\ref{eq:UL3})],  and  (iv) support-sufficient, magnetic-force-dominant models [outside the \bw relation line and below Equation~(\ref{eq:UL3})].
   In the models of class (i) [models (e), (f), (g), and (i)], the cloud collapses slowly, maintaining  spherical symmetry, and then a disk forms due to the rotation.
   In this type of evolution, the magnetic field hardly changes its initial direction.
   On the other hand, in the models of class (iii) [models (a), (b), and (c)], the cloud collapses along the rotation axis owing to the strong centrifugal force, and then a thin disk is formed in the early isothermal collapse phase.
   The magnetic field lines run along the disk plane, because the magnetic field lines are compressed together with a cloud in these models. 
   In the rotation-dominant models of classes (i) and (iii), the rotation axis maintains its initial direction because the magnetic braking is not so effective.
   On the other hand, in the magnetic-force-dominant models of classes (ii) and (iv), the rotation axis is inclined from the initial direction because the clouds with strong magnetic fields experience effective magnetic braking.
   The inclination of the rotation axis, $\theta_\Omega$ in class (iv) is greater than that in class (ii). 
   The direction of the magnetic field, however, tends to maintain its initial direction relative to the $z$-axis in classes (ii) and (iv) for a strong magnetic tension force.

\section{Discussion}
\subsection{Fragmentation of a Magnetized Rotating Cloud}
   Fragmentation is considered to be one of the mechanisms producing binary and multiple stars.
   We investigated fragmentation of a rotating magnetized cloud in Paper III for the case of $\vect{B_0} \parallel \vect{\Omega_0}$.
   We found fragmentation only in rotation-dominant clouds.
   This indicates that fragmentation is suppressed by the magnetic field, similar to the findings of \citet{hosking04} and \citet{ziegler05}.
   Fragmentation occurs via a global bar or ring mode and depends on the initial amplitude of the non-axisymmetric perturbation in support-sufficient rotation-dominant clouds. 
   That is, when the non-axisymmetric structure barely grows in the isothermal phase and the rotation rate reaches $\omega \gtrsim 0.2$ at the core formation epoch, an adiabatic core fragments via a ring (ring fragmentation).
   On the other hand, when the core is deformed to an elongated bar at the core formation epoch, the bar fragments into several cores (bar fragmentation).
  The parameter study in Paper III shows that ring fragmentation is seen in rotation-dominant models in the adiabatic phase [classes (i) and (iii)] and bar fragmentation is observed only in support-sufficient rotation-dominant models [class (iii)].
   Support-deficient and support-sufficient magnetic-dominant models [classes (ii) and (iv)] evolve into a single dense core without fragmentation owing to effective magnetic braking and a slow rotation.

   The results obtained in Paper III show that fragmentation patterns are dependent on the growth of a non-axisymmetric structure.
   The non-axisymmetric perturbation begins to grow after a thin disk is formed.
   The amplitude of the non-axisymmetric mode barely grows in support-deficient rotation-dominant models [class (i)], because the disk forms slowly, as shown in \S3.1.
   On the other hand, the non-axisymmetric structure grows sufficiently in support-sufficient rotation-dominant models [class (iii)], because the disk forms promptly.
   In class (iii), the patterns of fragmentation are dependent on the initial amplitude of the non-axisymmetric perturbation.
   Thus, a cloud fragments through a ring when the cloud has a small amount of initial non-axisymmetric perturbations, whereas the cloud fragments through a bar when it has a sufficient amount of initial non-axisymmetric perturbations.
These results apply to aligned rotator clouds.
Below we extend the study to the evolution of non-aligned rotator models.

   We did not add any explicit non-axisymmetric perturbations to the initial state of non-aligned rotator models.
  We did not find any rings for the rotation-dominant models of non-aligned rotators.
   In the rotation-dominant models, non-axisymmetry arises from the magnetic force in the case of $\vect{B_0} \nparallel \vect{\Omega_0}$.
 Since this non-axisymmetric perturbation from the magnetic force grows sufficiently in the isothermal phase,  bar fragmentation must occur  in non-aligned rotator models.

       As detailed previously, the axis-ratio ($\ar$), listed in Table~\ref{table:results}, arises from an anisotropic force due to the magnetic field or rotation.
   Model (c) has the greatest axis-ratio $\simeq 2.2$ of all models.
   The initial state of this cloud is outside the \bw relation line, and has the strongest magnetic field in rotation-dominant models (Figure~\ref{fig:17n}).
   In this cloud, a disk is formed perpendicular to the rotation axis and  the magnetic field parallel to the disk surface is greatly amplified.
This induces a bar structure along the magnetic field line on the disk mid-plane, as shown in \S 3.2 (e.g. model C45).
   Although this bar does not fragment in this study, such a bar structure suggests the possibility of fragmentation in the adiabatic phase (Paper III) if the calculation is continued.
   Even if the bar does not fragment in the case of there being no explicit non-axisymmetric perturbations, we expect bar fragmentation if an initial explicit perturbation is added.
  We confirmed that bar fragmentation occurs in model (c) when we added a 10\% non-axisymmetric density perturbation to the initial state.
   As a result, the anisotropy arising from magnetic and centrifugal forces in non-aligned rotator models promotes bar fragmentation and suppresses ring fragmentation.

\subsection{Comparison with Previous Works}
   Several studies have examined the gravitational collapse of molecular cloud cores and star formation using three-dimensional MHD calculations  \citep[][MT04; Papers I, II and III]{dorfi82, dorfi89,boss02,hosking04,ziegler05, banerjee06}.
   Except for those of \citet{dorfi82,dorfi89} and MT04, these studies assume that the magnetic field lines are initially parallel to the rotation axis.
   The evolutions of non-aligned rotator models are almost the same as those of aligned rotator models for support-deficient models, while the evolutions are completely different for support-sufficient models.
   For example, the direction of the magnetic field continues to move away from the rotation axis in a rapidly rotating cloud.
  The directions of the magnetic field and rotation are completely different after disk formation in this case.
  Comparing magnetic fields parallel and perpendicular to the rotation vector shows that a perpendicular  magnetic field transfers angular momentum more effectively than a parallel field \citep{mouschovias79}.
 In other words, magnetic braking is more effective in non-aligned rotator models than in aligned rotator models.   
   This strong magnetic braking may be the solution of the angular momentum problem, in which the specific angular momentum of a parent cloud is much larger than that of a new-born star.

\subsection{Comparison with Observation}
\begin{figure}  
\epsscale{1.0}
\plotone{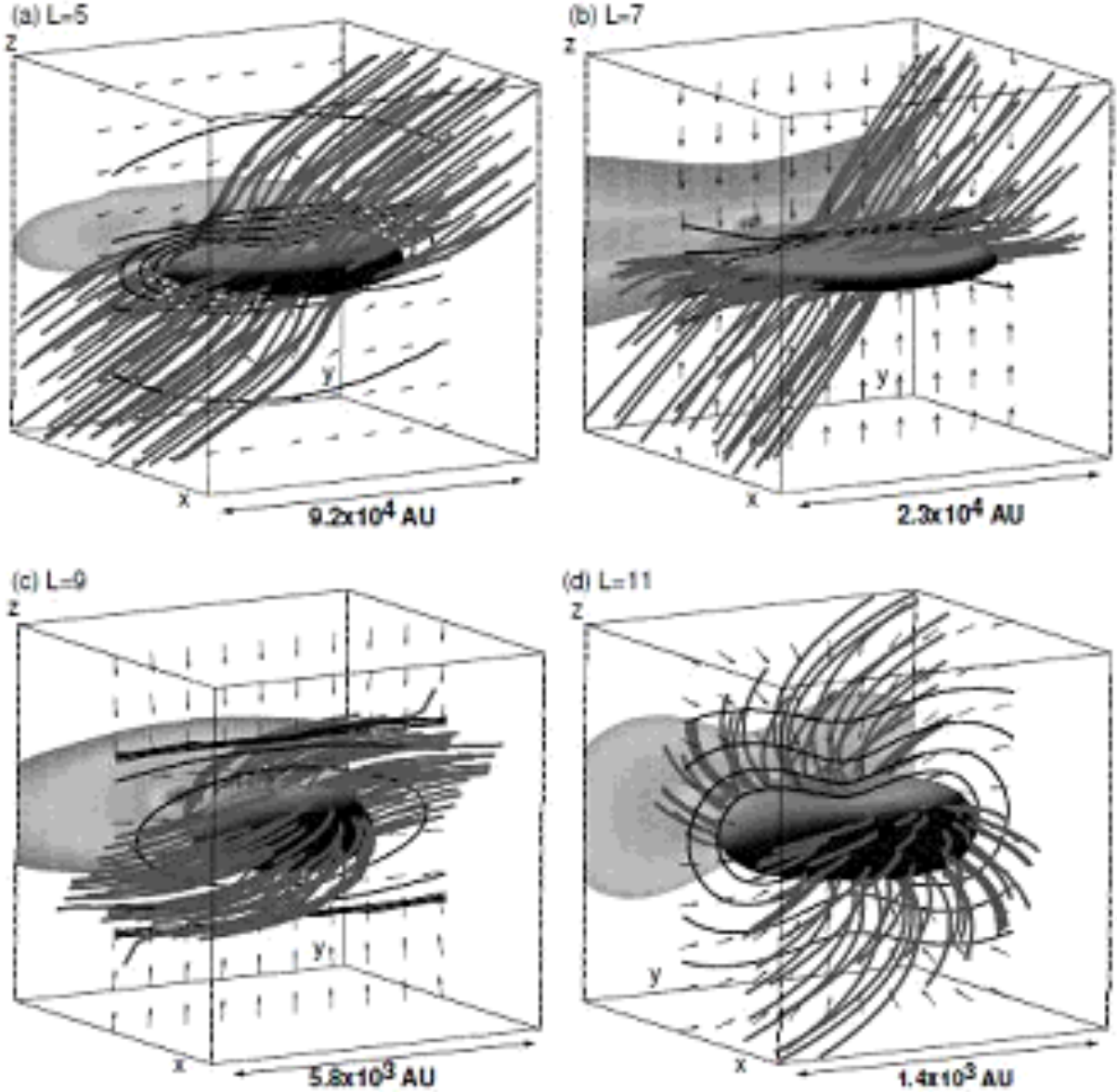}
\caption{
Views of model C30.
Each panel shows the same epoch, $t= 1.52 \times 10^6$ yr ($n_c = 1.5 \times 10^9 \cm$), but for different box scales, (a) $9.2 \times 10^4$ AU, (b) $2.3 \times 10^4$ AU, (c) $5.8 \times 10^3$ AU, and (d) $1.4 \times 10^3$ AU.
Isodensity surfaces are drawn at the densities (a) $n_c \ge 1.3 \times 10^4 \cm$, (b) $n_c \ge 5 \times 10^4$, (c) $n_c \ge 5\times 10^6$, and (d) $n_c \ge 5\times 10^7\cm$.
The density distribution on the  $y$ = 0 plane is projected on to the boundary by a false color plot.
The other representatives are the same as those in Figure~\ref{fig:1}.
}
\label{fig:13n}
\end{figure}

\begin{figure}  
\plotone{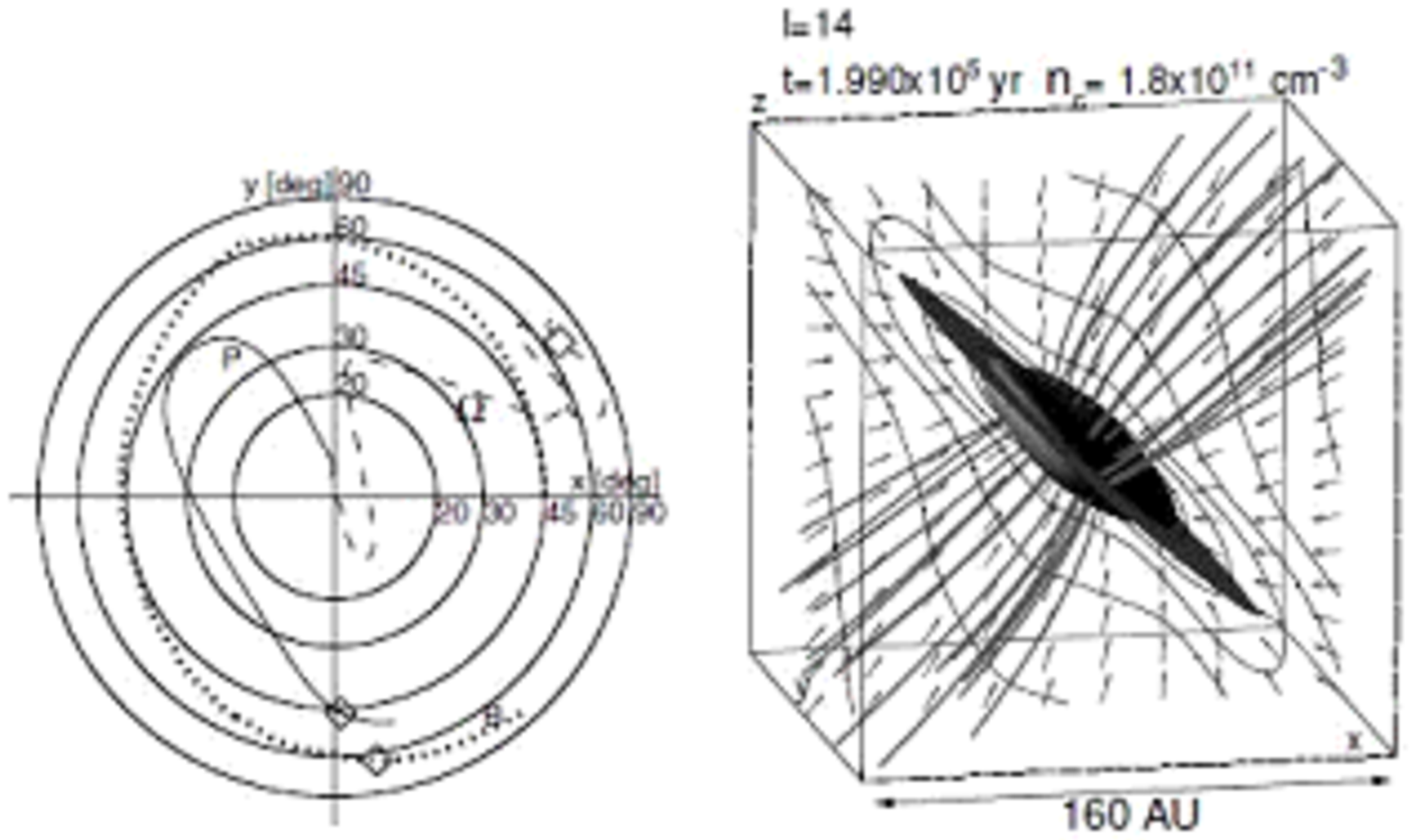}
\caption{
As Figure~\ref{fig:5} for model D45
at $n_c = 1.8 \times 10^{11} \cm$ (right).
}
\label{fig:14n}
\end{figure}

\begin{figure}  
\vspace{-1.5cm}
\plotone{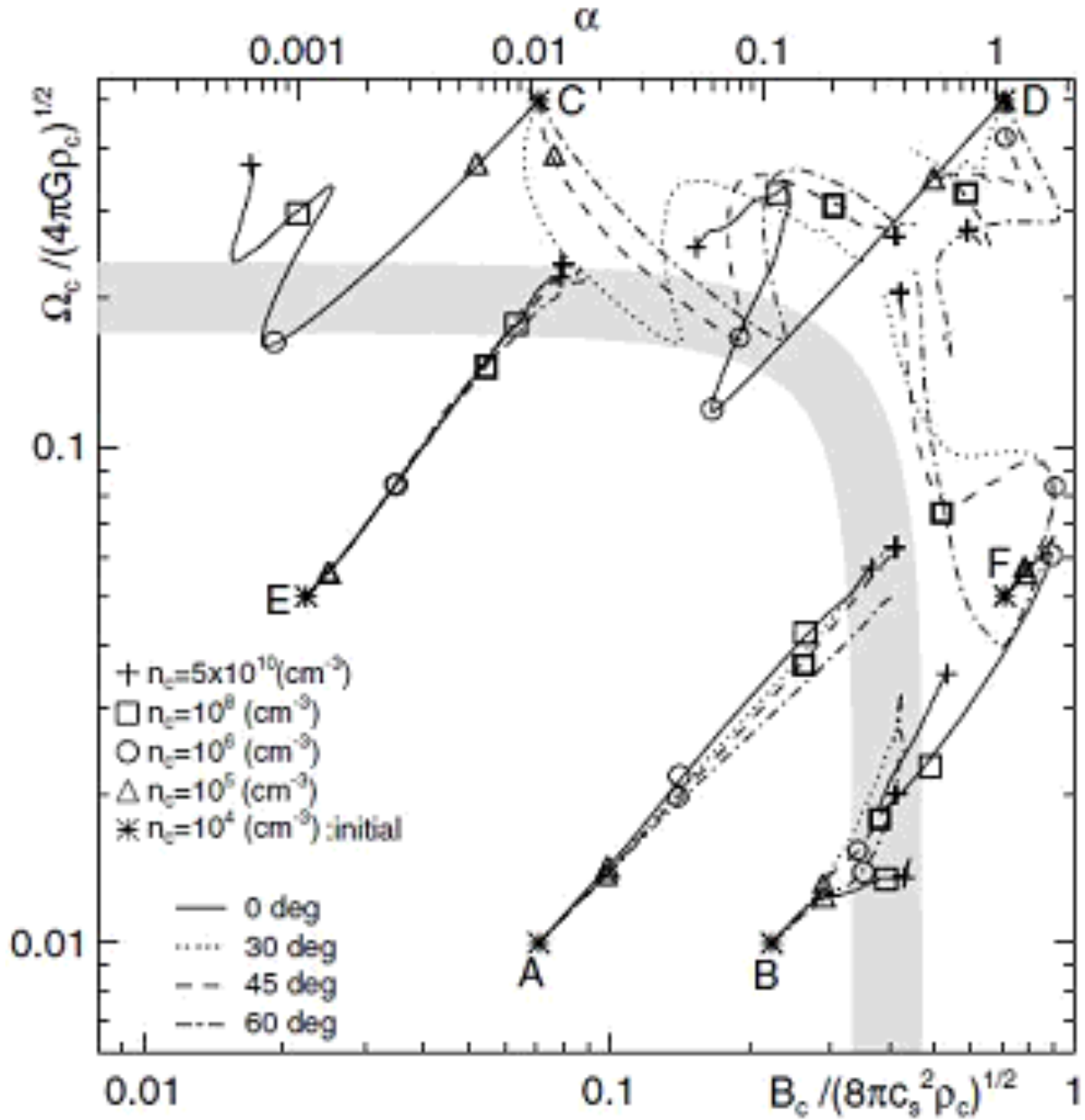}
\vspace{-1cm}
\caption{
Evolution of magnetic flux density and angular velocity at the cloud center.
The $x$-axis indicates the square root of the magnetic pressure [$B_c\,/ (8\pi)^{1/2}$] divided  by the square root of the thermal pressure [$ (c_s^2 \rho_c)^{1/2}$]. The $y$-axis represents the angular speed ($\Omega_c$) divided by the free-fall rate [$(4 \pi G \rho_c)^{1/2}$].
Here, we use the definitions $B_c = (B_{x,c}^2 + B_{y,c}^2 + B_{z,c}^2)^{1/2}$ and $\Omega_c = (\Omega_{x,c}^2 + \Omega_{y,c}^2 + \Omega_{z,c}^2)^{1/2}$, where the suffix $c$ indicates the value at the center.
The upper $x$-axis indicates the value of $\alpha$.
The symbols $*$, $\vartriangle$, $\circ$, $\square$, and $+$ represent the magnetic field and the angular velocity at $n_c = 5 \times 10^4 \cm$ (initial state),  $10^5 \cm$, $ 10^6 \cm$, $ 10^8 \cm$, and $5\times 10^{10} \cm$,
respectively.
Each line denotes the evolution path from the initial state ($n_{\rm c,0} = 5\times 10^4 \cm$) to the end of the isothermal phase ($n_{\rm c} = 5 \times 10^{10} \cm$).
The characters A, B, C, D, and E denote the group names shown in Table~\ref{table:init}.
The different lines indicate different initial angles,  $\theta_0 = 0 \degr$ (solid line), $30\degr$ (dotted line), $45\degr$ (broken line), and $60\degr$ (dash dotted line).
The thick grey band denotes the magnetic flux--spin relation $\Omega_c^2/[(0.2)^2 \times 4\pi G \rho_c] + B_{c}^2/[(0.36)^2 \times 8\pi c_s^2 \rho_c] =1$ [see Equation~(\ref{eq:UL2})].
}
\label{fig:15n}
\end{figure}

\begin{figure}  
\plotone{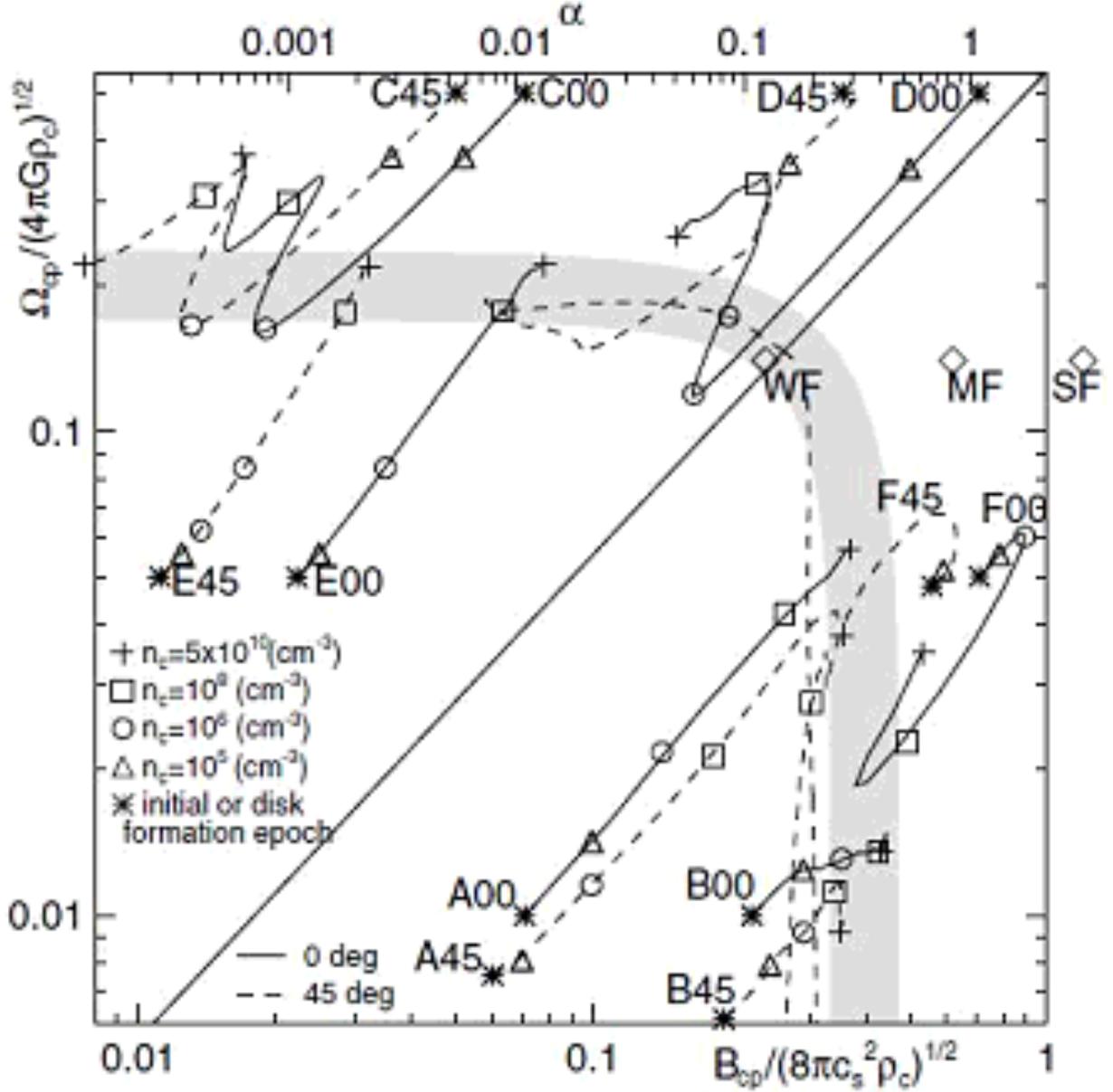}
\caption{
Evolution of magnetic flux density and angular velocity parallel to the disk normal.
The axes and symbols are the same as for Figure~\ref{fig:15n}.
Here, $B_{cp}$ and $\Omega_{cp}$ are the magnetic field and angular velocity parallel to the disk normal [Equations~(\ref{eq:bp}) and (\ref{eq:wp})].
Models with $\theta_0$ = $0\degr$ and $45\degr$ are plotted.
The symbols $\diamond$ show the initial states of the models WF, MF, and SF calculated in MT04 (see Table~\ref{table:init}).
}
\label{fig:16n}
\end{figure}

\begin{figure}  
\plotone{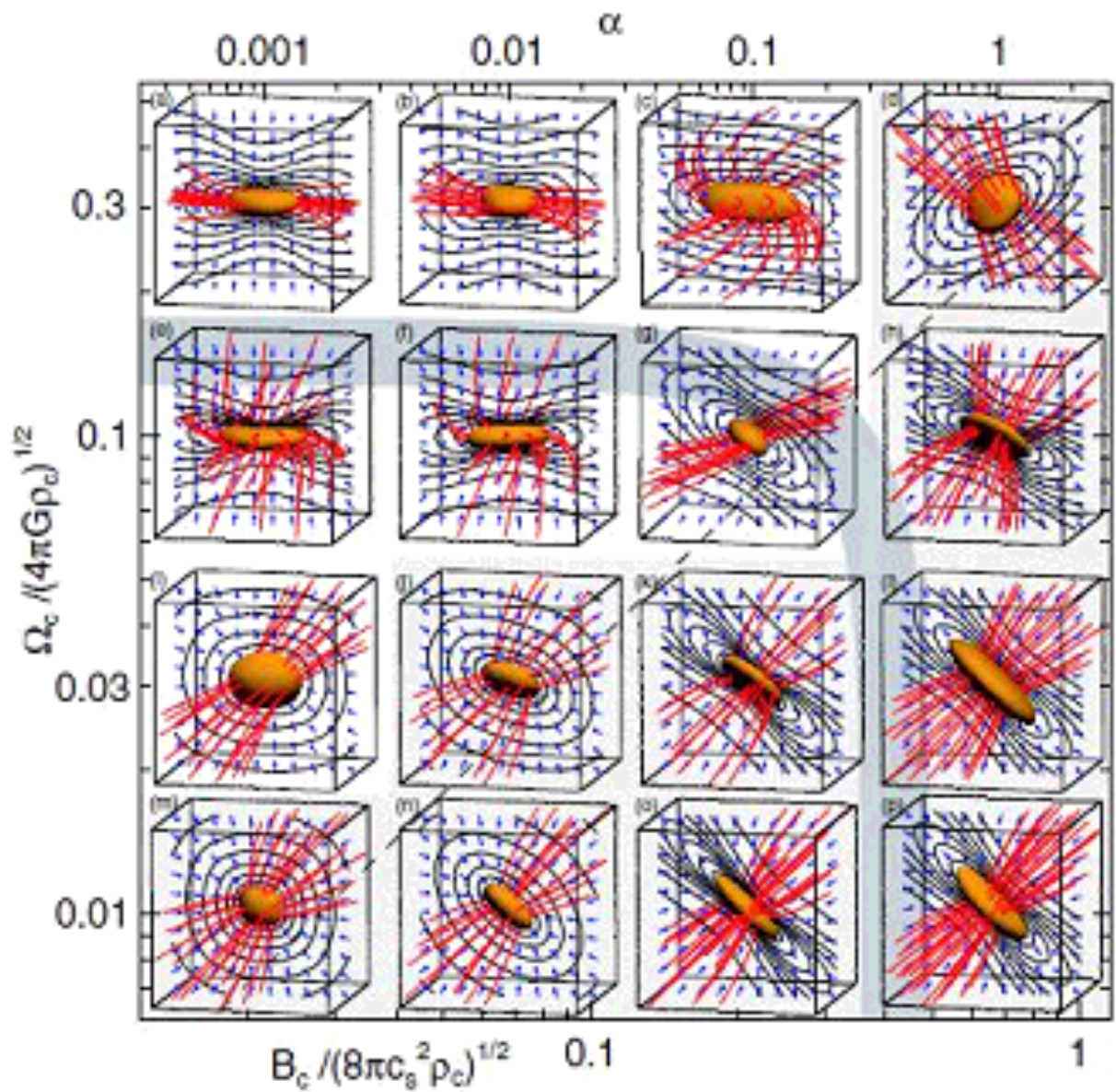}
\caption{
Snapshots of a disk and magnetic field lines at $n_c = 5\times 10^{10} \cm$.
The snapshots are displayed according to the initial values of $\Omega_c$ and $B_c$.
The axes and curve are the same as that of Figure~\ref{fig:15n}.
The structure of the high-density region ($n > 0.1 n_c$\,; isodensity surface), density contours (contour lines), velocity vectors (arrows), and magnetic field lines (streamlines) are plotted in the panels 
(a)--(p).
The shaded part indicates the region in which a magnetic-dominant disk is formed.
The broken line denotes the border between the rotation- and the magnetic-force-dominant disks, $\Omega /B   = 0.39 G^{1/2}\, c _s^{-1}$ [Equation~(\ref{eq:UL3})].
}
\label{fig:17n}
\end{figure}

    The magnetic field strengths and directions have been observed for many clouds.  
    It is believed that there is no correlation between the direction of the magnetic field and the large-scale cloud shape \citep{goodman93,tamura95,word00}.
    Recently, the directions of the magnetic field have been observed in both large (cloud) and small (prestellar core) scales in the same target.
    The small-scale polarization pattern of the W51 molecular cloud observed by BIMA \citep[Berkeley-Illinois-Maryland Association;][]{lai01} coincides with the large-scale polarization pattern observed by SCUBA \citep{chrysostomou02}. 
    Also, the directions of the magnetic field were found to be the same in both large- and small-scale observations of the DR21 cloud \citep{lai03}. 
    However, some observations have given contrasting findings.
    Although the average polarization angle in the MMS6 core in the OMC-3 region of the Orion A cloud \citep{matthews05} coincides with the large-scale polarization angle observed by SCUBA \citep{houde04}, the polarization angle changes systematically across the core.
    An observation of the Barnard 1 cloud in Perseus reveals that three of the four cores exhibit different mean field directions than that of the ambient cloud \citep{matthews02,matthews05}.   
    These trends of OMC-3 and the Barnard 1 cloud agree well with the results for non-aligned rotator models in group C in \S 3.1.
   The direction of a small-scale magnetic field can be different from the large-scale field in models C30, C45, and C60 (support-sufficient, rotation-dominant models).
   As shown in Figure~\ref{fig:13n} for model C30, although the magnetic field maintains its initial direction $\theta_{B} \simeq 30 \degr$ outside the high-density core, inside the high-density region the magnetic field lines are perpendicular to the rotation ($z$-) axis ($\theta_{B} \simeq 90 \degr$).
   Thus, the direction of the magnetic field varies for different spatial scales.
   In this model, the angle between the large-scale and small-scale magnetic fields is $85 \degr$.
   On the other hand, the magnetic field lines have the same direction in both large- and small-scales in W51 and DR21.
   These clouds correspond to groups A, B, and E, in which the magnetic field hardly changes its direction in the isothermal phase.
  These clouds are expected to have a slow rotation rate.
   These findings show that the direction of the magnetic field can change only in support-sufficient rotation-dominant clouds.

  Recently, an hour-glass-shaped magnetic field has been found in a dynamically contracting core around the binary protostellar system NGC 1333 IRAS 4A \citep{girart06}.
  Two outflows were also observed in this region and associated with each protostar of the protobinary system \citep{choi05}.
  However, the direction of the magnetic field does not coincide with the outflow axis \citep{girart06}.
  From our previous study, fragmentation (or binary formation) appears only in rotation-dominated clouds (Paper III).
  In these clouds, the magnetic field tends to be aligned along the direction perpendicular to the rotation axis, as shown in \S 3.2.1,
  and therefore the direction of the magnetic field in a dense core does not coincide with that of the large-scale field.
  Thus, the observed miss-aligned outflow indicates that a binary is being formed from a rotation-dominated cloud, because the outflows are driven along the local magnetic field (MT04).

\acknowledgments
Our numerical calculations were carried out with a Fujitsu VPP5000 
at the Astronomical Data Analysis Center 
of the National Astronomical Observatory of Japan.
This work was supported partially by Grants-in-Aid from MEXT (16077202 [MM], 17340059 [TM, KT], 15340062, 14540233 [KT], 16740115 [TM]).

\clearpage

\begin{table}   
\setlength{\tabcolsep}{1pt}
\caption{Parameters and initial conditions for typical models}
\label{table:init}
\begin{center}
\begin{tabular}{ccccccccccccccccccc}
\hline
 \multicolumn{2}{c}
Group & Model & $\alpha$  & $\omega$ & $\theta_0$  & $f$  &$\alpha_0$ &$\beta_0$ &$\gamma_0$ & $n_{0}$$^a$ & $B_0$$^b$ & $\Omega_0$$^c$  &  $M^d$ \ 
 \\
\hline 
&A& {\footnotesize A00, A30, A45, A60}& 0.01& 0.01& {\footnotesize(0$\degr$, 30$\degr$, 45$\degr$, 60$\degr$)}& 1.05&0.70&3.29$\times10^{-4}$ &1.34$\times10^{-2}$ 
&5.25 & 3.23 & 1.38 & 6.41  \\

&B& {\footnotesize B00, B30, B45, B60}&  0.1& 0.01& {\footnotesize(0$\degr$, 30$\degr$, 45$\degr$, 60$\degr$)}& 1.05&0.70&3.29$\times10^{-4}$ &0.134 
&5.25 &10.2 & 1.38& 6.41  \\

&C& {\footnotesize C00, C30, C45, C60}& 0.01& 0.5 & {\footnotesize(0$\degr$, 30$\degr$, 45$\degr$, 60$\degr$)}& 5.0 &0.168&0.823&3.22$\times10^{-3}$ 
&25&6.59 & 141 & 26.7  \\

&D& {\footnotesize D00, D30, D45, D60}& 1   & 0.5 & {\footnotesize(0$\degr$, 30$\degr$, 45$\degr$, 60$\degr$)}& 5.0 &0.168&0.823&0.32
&25&65.9& 141 & 26.7 \\

&E& {\footnotesize E00, E30, E45, E60}& $10^{-3}$& 0.05 & {\footnotesize(0$\degr$, 30$\degr$, 45$\degr$, 60$\degr$)}& 5.0&0.168&0.82$\times10^{-3}$ &3.22$\times 10^{-4}$  
&25&2.08 & 14.1 & 26.7  \\

&F& {\footnotesize F00, F30, F45, F60}& 1   & 0.05 & {\footnotesize(0$\degr$, 30$\degr$, 45$\degr$, 60$\degr$)}& 5.0 &0.168 &0.82$\times10^{-3}$&0.32
&25&65.9& 14.1 & 26.7 \\

\hline
& & SF$^e$&3.04&0.14 & {\footnotesize (00$\degr$, 45$\degr$, 90$\degr$) }       & 1.68& 0.5 & 0.02 & 2.88 & 2.61 & 37.1 & 7.11 & 6.13  \\
& & MF$^e$&0.76&0.14 &{\footnotesize  (45$\degr$, 70$\degr$, 80$\degr$)} & 1.68& 0.5 & 0.02 & 0.72 & 2.61 & 18.6 & 7.11 & 6.13 \\
& & WF$^e$&0.12&0.14 &{\footnotesize (00$\degr$, 45$\degr$, 90$\degr$)       }& 1.68& 0.5 & 0.02 & 0.12 & 2.61 & 7.42 & 7.11 & 6.13  \\
\hline
\end{tabular}
\\{\small
$^a$ $n_{0} (10^4 \times \cm)$, $^b$ $B_0$ ($\mu \rm{G}$), $^c$ $\Omega_0$ ($10^{-7}$ yr$^{-1}$), $^d$ $M$ ($\msun$) , $^e$ models calculated by \cite{matsumoto04}
}
\end{center}
\end{table}

\clearpage
\begin{table}   
\renewcommand{\arraystretch}{0.8}
\setlength{\tabcolsep}{5pt}
\vspace{-2cm}
\caption{Calculation results at the core formation epoch} 
\label{table:results}
\begin{center}
\begin{tabular}{ccccc|cccccccccccccc}
\hline
 Model  & $\alpha$  & $\omega$ & $\theta_0$ & $f$&{\footnotesize($\theta_B$, $\theta_\Omega$, 
$\theta_P$) } & {\footnotesize($\phi_B$, $\phi_\Omega$, $\phi_P$)} & {\footnotesize ($\psi_{B \Omega}$, $\psi_{B P}$, $\psi_{\Omega P}$)}& {\small $B$/$\Omega$}
\tablenotemark{a}
 &$\ar$ \\
\hline
A00 &  0.01   & 0.01 & 00& 1.05&(0, 0, 0)& (0, 0, 0)  & (0, 0, 0)& 
--- &0\\
A30 &  0.01   & 0.01 & 30& 1.05&(31, 1, 30)& (24, 22, 20)  & (30, 2, 29)& 
$B$ & $1.2 \times 10^{-3}$\\
A45 &  0.01   & 0.01 & 45& 1.05&(46, 2, 44)& (24, 23, 20)  & (44, 4, 42)& 
$B$ &$7.1 \times 10^{-3}$\\
A60 &  0.01   & 0.01 & 60& 1.05&(61, 2, 58)& (24, 11, 20)  & (59, 4, 56)& 
$B$ &$1.5 \times 10^{-2}$\\
\hline
B00 &  0.1   & 0.01 & 00& 1.05&(0, 0, 0)& (0, 0, 0)  & (0, 0, 0)& 
--- &0\\
B30 &  0.1   & 0.01 & 30& 1.05&(33, 32, 33)& (22, 66, 14)  & (24, 4, 27)& 
$B$ &$8.0 \times 10^{-3}$\\
B45 &  0.1   & 0.01 & 45& 1.05&(46, 7, 45)& (23, 14, 15)  & (39, 5, 38)& 
$B$ &$2.5 \times 10^{-3}$\\
B60 &  0.1   & 0.01 & 60& 1.05&(58, 52, 57)& (24, 298, 15)  & (68, 8, 61)& 
$B$ &$1.3 \times 10^{-2}$\\
\hline
C00 &  0.01   & 0.5 & 00& 5&(0, 0, 0)& (0, 0, 0)  & (0, 0, 0)& 
--- &0\\
C30 &  0.01   & 0.5 & 30& 5&(88, 1, 2)& (302, 273, 96)  & (88, 90, 3)& 
$\Omega$ &0.43\\
C45 &  0.01   & 0.5 & 45& 5&(90, 1, 2)& (302, 263, 86)  & (88, 89, 3)& 
$\Omega$ &0.61\\
C60 &  0.01   & 0.5 & 60& 5&(89, 1, 2)& (302, 257, 84)  & (89, 89, 3)& 
$\Omega$ &0.68\\
\hline
D00 &  1   & 0.5 & 00& 5&(0, 0, 0)& (0, 0, 0)  & (0, 0, 0)& 
--- &0\\
D30 &  1   & 0.5 & 30& 5&(60, 13, 14)& (243, 8, 34)  & (67, 72, 6)& 
$\Omega$ &0.19\\
D45 &  1   & 0.5 & 45& 5&(51, 63, 49)& (283, 39, 275)  & (88, 6, 85)& 
$B$ &0.21\\
D60 &  1   & 0.5 & 60& 5&(35, 26, 51)& (337, 51, 13)  & (36, 29, 33)& 
$B$ &0.32\\
\hline
E00 &  0.001& 0.05 & 00& 5&(0, 0, 0)& (0, 0, 0)  & (0, 0, 0)& 
--- &0\\
E30 &  0.001& 0.05 & 30& 5&(43, 1, 1)& (50, 80, 38)  & (43, 42, 1)& 
$\Omega$ &$3.4\times 10^{-3}$\\
E45 &  0.001& 0.05 & 45& 5&(59, 1, 1)& (50, 110, 47)  & (58, 58, 1)& 
$\Omega$ &$6.8\times 10^{-3}$\\
E60 &  0.001& 0.05 & 60& 5&(71, 1, 1)& (90, 142, 49)  & (71, 70, 1)& 
$\Omega$ &$1.3\times10^{-2}$\\
\hline 
F00 &  1& 0.05 & 00& 5&(0, 0, 0)& (0, 0, 0)  & (0, 0, 0)& 
--- &0\\

F30 &  1& 0.05 & 30& 5&(35, 9, 30)& (51, 13, 24)  & (28, 15, 21)& 
$B$ &0.21\\
F45 &  1& 0.05 & 45& 5&(51, 10, 44)& (51, 354, 24)  & (46, 20, 35)& 
$B$ &0.40\\
F60 &  1& 0.05 & 60& 5&(65, 12, 58)& (51, 336, 24)  & (62, 24, 50)& 
$B$ &0.40\\
\hline 

(a) & 0.001& 0.3 & 45& 1.68&(88, 0, 0)& (189, 308, 224)   & (88, 88, 0)& 
$\Omega$ &$5.9\times 10^{-2}$\\
(b) & 0.01 & 0.3 & 45& 1.68&(86, 2, 0)& (294, 269, 87)  & (85, 87, 2)& 
$\Omega$ &0.51\\
(c) & 0.1  & 0.3 & 45& 1.68&(82, 15, 8)& (226, 356, 167)   & (89, 78, 23)& 
$\Omega$ &2.2\\
(d) &  1   & 0.3 & 45& 1.68&(48, 46, 42)& (256, 32, 231)  & (85, 18, 87)& 
$B$ &0.17\\
\hline
(e) & 0.001& 0.1 & 45& 1.68&(83, 0, 0)& (120, 42, 127)  & (83, 83, 1)& 
$\Omega$ &$1.7\times 10^{-2}$\\

(f) & 0.01 & 0.1 & 45& 1.68&(83, 3, 4)& (125, 4, 126)  & (85, 79, 6)& 
$\Omega$ &0.18\\
(g) & 0.1  & 0.1 & 45& 1.68&(78, 16, 37)& (146, 347, 109)   & (87, 50, 48)& 
$\Omega$ &0.44\\
(h) &  1   & 0.1 & 45& 1.68&(43, 32, 42)& (63, 68, 60)& (12, 2, 11)& 
$B$ &$5.2\times 10^{-2}$\\
\hline
(i) & 0.001& 0.03& 45& 1.68&(53, 0, 4)& (51, 81, 49) & (53, 49, 4)& 
$\Omega$ &$2.1\times 10^{-2}$\\
(j) & 0.01 & 0.03& 45& 1.68&(54, 2, 30)& (53, 25, 45)& (52, 25, 28)& 
$B$ & $9.1\times 10^{-2}$\\
(k) & 0.1  & 0.03& 45& 1.68&(58, 5, 45)& (70, 342, 35) & (58, 30, 40)& 
$B$ &$6.0\times 10^{-2}$\\
(l) &  1   & 0.03& 45& 1.68&(45, 26, 45)& (20, 16, 19)& (19, 1, 19)& 
$B$ &$5.3\times 10^{-2}$\\
\hline
(m) & 0.001& 0.01& 45& 1.68&(46, 0, 29)& (18, 2, 16)& (46, 17, 29)&  
$B$ & $1.1\times 10^{-2}$\\
(n) & 0.01 & 0.01& 45& 1.68&(46, 5, 43)& (19, 21, 16)& (41, 4, 39)& 
$B$ &$6.0\times 10^{-3}$\\
(o) & 0.1  & 0.01& 45& 1.68&(45, 8, 45)& (29, 4, 12)& (38, 12, 37)& 
$B$ &$5.8\times 10^{-3}$\\
(p) &  1   & 0.01& 45& 1.68&(45, 29, 45)& (6, 5, 6)& (16, 0, 16)& 
$B$ &$2.8\times 10^{-2}$\\
\hline
\end{tabular}
\end{center}
\label{table:2}
\tablenotetext{a}{
The dominant forces for disk formation. They are determined by the loci of the magnetic field, the rotation axis, and disk normal for the isothermal phase.
}
\end{table}


\newpage

\clearpage

\end{document}